\documentclass[tighten, twocolumn, trackchanges, 12pt]{aastex62}
\def\apjl{{ApJ}}                
\def\apjs{{ApJS}}               
\def\ao{{Appl.~Opt.}}           
\def\aap{{A\&A}}                

\usepackage{listings}
\usepackage{color}
\usepackage{graphicx,times}
\usepackage{natbib}
\usepackage{amssymb}
\usepackage{newtxtext}
\usepackage[T1]{fontenc}
\usepackage{ae,aecompl}
\usepackage{threeparttable}
\usepackage{ulem}
\usepackage{txfonts}
\usepackage{amssymb}	

\def\kms  {km~s$^{-1}$}

\bibpunct{(}{)}{;}{a}{}{,}
\usepackage{hyperref}
\usepackage{longtable}
\usepackage{booktabs}
\definecolor{dkgreen}{rgb}{0,0.6,0}
\definecolor{gray}{rgb}{0.5,0.5,0.5}
\definecolor{mauve}{rgb}{0.58,0,0.82}
\definecolor{golden}{rgb}{0.86,0.65,0.01}

\usepackage{CJK}
\usepackage{graphicx}
\usepackage{hyperref}  
\begin{document}
\begin{CJK*}{UTF8}{gbsn}
\received{28 March}
\revised{9 June}
\accepted{23 June}

\title[]{A Blind All-sky Search for Star Clusters in Gaia EDR3: 886 Clusters within 1.2~kpc of the Sun}
\correspondingauthor{Zhihong He}
\email{hezh@cwnu.edu.cn}
\author[0000-0002-6989-8192]{Zhihong He (何治宏)}
\affil{School of Physics and Astronomy, China West Normal University, No. 1 Shida Road, Nanchong 637002, China }
\author{Kun Wang (王坤)}
\affil{School of Physics and Astronomy, China West Normal University, No. 1 Shida Road, Nanchong 637002, China }
\author{Yangping Luo (罗杨平)}
\affil{School of Physics and Astronomy, China West Normal University, No. 1 Shida Road, Nanchong 637002, China }
\author{Jing Li (李静)}
\affil{School of Physics and Astronomy, China West Normal University, No. 1 Shida Road, Nanchong 637002, China }
\author{Xiaochen Liu (刘效臣)}
\affil{School of Physics and Astronomy, China West Normal University, No. 1 Shida Road, Nanchong 637002, China }
\author{Qingquan Jiang (蒋青权)}
\affil{School of Physics and Astronomy, China West Normal University, No. 1 Shida Road, Nanchong 637002, China }
\vspace{10pt}

\begin{abstract}
Although previous searches for star clusters have been very successful, many clusters are likely still omitted, especially at high Galactic latitude regions. In this work, based on the astrometry of Gaia EDR3, we searched nearby ($\varpi$ > 0.8~mas) all-sky regions, obtaining 886 star clusters, of which 270 candidates have not been cataloged before. At the same time, we have presented the physical parameters of the clusters by fitting theoretical isochrones to their optical magnitudes. More halo members and expanding structures in many star clusters were also found. Most of the new objects are young clusters that are less than 100 million years old. 
Our work greatly increased the sample size and physical parameters of star clusters in the solar neighborhood, in particular, 46 clusters are newly found with |b| > 20$^{\circ}$, which represents an increase of nearly three fold of cluster numbers at high Galactic latitude regions.
The cluster parameters and member stars are available at CDS via \url{https://cdsarc.u-strasbg.fr/ftp/vizier.submit//hezh22b/}, and the cluster figure sets are available via \url{https://doi.org/10.12149/101133}.
\end{abstract}

\section{Introduction}\label{sec:intro}

For decades, open clusters (hereafter OCs) have been an important tracer to study the  structural and chemical evolution of the Galactic disk~\citep[][]{Friel95,xu18,Spina22}, as well as an important observational laboratory for stellar evolution~\citep{Lada03,araa2010}. Limited by Galactic dust extinction and field star contamination, the identification of cluster members and the establishment of complete cluster samples are a difficult task. However, Gaia satellite~\citep[][]{Gaia16} observations have yielded a large amount of stellar data with accurate astrometric measurements~\citep{Gaia18-Brown,gaia2021}, and many studies based on these data have brought a huge increase in the number of identified star clusters~\citep[][]{Cantat22}. 

In space near the Solar System, the influence of distance and extinction is smaller than for much further objects. As such, studies of objects near to the Solar System can provide a unique opportunity to obtain more complete cluster samples and richer member information. Based on the Gaia Second Data Release (DR2), since 2018 a large number of nearby star clusters have been identified~\citep[][]{CG18}, and more new star clusters have been found~\citep[e.g.][]{Liu19,Sim19,Castro20,he21,he22}. 
The members of known clusters are constantly updated~\citep[e.g.][]{Gao18,CG20_0,Monteiro20,Jaehnig21,Tarricq22}, and many physical and structural characteristics of these clusters have also been presented~\citep[e.g.][]{Kounkel19,Cantat19_exp,Zhong20,Tian20,Pang21}.

Different from asterisms or field stars along a given line of sight, the member stars of a cluster have the same characteristics in terms of spatial and kinematic locations; this allows stellar aggregates to be found by clustering based on astrometric data~\citep{Castro18,Liu19}. Up to now, the majority of new clusters were detected through the DBSCAN clustering algorithm~\citep[e.g.][]{Castro20,castro22,he21,he22} and its improved method HDBSCAN~\citep[e.g.][]{Kounkel19,Hunt21}. DBSCAN has been demonstrated to be a very efficient blind search method in big-data approaches~\citep[][]{gao14,Cantat22}.

However, all of above automatic clustering studies focused on the Galactic plane, specifically Galactic latitudes of |b| < 20$^{\circ}$ to 30$^{\circ}$, and the higher Galactic latitude regions have only been manually searched~\citep[][]{Sim19}. Since the Galactic altitude of OCs can reach |Z| = 200 to 400 pc~\citep{CG18}, high Galactic latitude areas within $\sim$1 kpc of the Solar System may not have been fully explored.

In our previous studies~\citep[][hereafter H21, H22, respectively]{he21,he22}, we used Gaia DR2 data to systematically search for new OCs in the Galactic disk. We revisited most of them using Gaia Early Data Release 3 (EDR3), which resulted in a total about 600 candidates with individual membership information for cluster stars. However, owing to the limited size of the searching grid (2 to 3 degrees), and an incomplete clustering parameter setting in the program (H22), clusters within $\sim$0.5 kpc was absent from these searches.

In this work, we aimed to identify cluster-like objects in nearby space ($\varpi$ > 0.8~mas) and determine their fundamental parameters. We applied the DBSCAN clustering method to all-sky regions, and expand our search grid to 12 to 18 degrees, hoping to considerably increase the number of clusters in the solar vicinity and built a homogeneous set of cluster parameters and collect more cluster members. 

The remainder of this paper is organized as follows. In Section~2, we describe the data preparation, including specifics of the data criteria, and the selection of data subsets. We describe the method that was applied to detect clusters in Section~3, and introduce how we determine the cluster ages, and extinctions via isochrone fits. In Section~4, we present our main results and discussion: we estimate the statistical properties of the cluster samples, cross match our results with published catalogs, classify the new cluster samples, and discuss the distribution of clusters in the solar vicinity. Finally, we present our conclusions in Section~5.

\section{DATA  PREPARATION}\label{sec:data}
For our study, the basic stellar data were taken from Gaia EDR3~\citep{gaia2021}, which contains parallaxes, photometry (G, G$_{BP}$, G$_{RP}$), and positions/proper motions from the International Celestial Reference System for about 1.5 billion objects down to G$\sim$21 mag. For about 7.2 million stars, the catalog contains radial velocities from Gaia DR2~\citep{Gaia18-Brown}. 

Nearby star clusters, especially some extended ones, have a wide projected distribution in the sky~\citep[][]{Kounkel19,Cantat19_exp,Sim19}, requiring that the search range must be large enough. However, at the same time, such large search areas will generate a large amount of data, which will increase the time consumed during each search and cause contamination to the procedure. Considering the above two points, faint and distinct objects were excluded: only stars with a G-band magnitude lower than 18 mag were used, and we also adopted parallax cuts of $\varpi$ > 1.8~mas (for subset~1) and 0.6 mas < $\varpi$ < 2.2~mas (for subset~2). The data sets were shifted with 0.2~mas to make sure that cluster members near the parallax cuts in each subset could be completely detected. 

After applying these criteria, we extracted Gaia EDR3 astrometry subsets for all selected stars in a 18$^{\circ}$ × 18$^{\circ}$ (for subset 1) or 12$^{\circ}$ × 12$^{\circ}$ (for subset 2) sized grid around each reference point, using intervals of 9$^{\circ}$ (for subset 1) or 6$^{\circ}$ (for subset 2). This approach ensured that as many members of a cluster as possible could be detected in a given grid. At the same time, we also considered the projection effect in the longitude direction with a factor of $\sec$~$b$, and extended the longitude around $l \sim$ 0$^{\circ}$/360$^{\circ}$.

As described by H21 and H22, for each stellar datum, we converted the position and proper motion to a projected linear distance and linear velocity relative to a reference center as: 
\begin{equation}\label{eq1}
(d_{l^*},d_b,v_{\alpha^*},v_{\delta})=
 (d\cdot\sin\theta_{l}\cdot\cos b,d\cdot\sin\theta_b,d\cdot\mu_{\alpha^*},d\cdot\mu_{\delta})
\end{equation}
where $d$ is taken to be the inverse of the parallax, and $\theta_{l}$ and $\theta_{b}$ are the angular sizes from the star to the reference center, respectively. Each element in each vector was standardized using a cluster median dispersion value from~\citet[][hereafter CG20]{cg20arm}. 
\section{Method}\label{sec:method}

\subsection{Clustering}\label{clustering}

The DBSCAN algorithm was used to count the number of data points within radius $\epsilon$ in phase space $(d_{l^*},d_b,v_{\alpha^*},v_{\delta},\varpi)$. Inspired by ~\citet[][]{gao14,gao17,Castro18}, we adopted an improved method, originally used by H21, in the selection of $\epsilon$. First, we calculated the $k_{th}$ nearest neighbor distance ($kNND$) to each point in the data set. Then, a bimodal Gaussian curve was fitted to the $kNND$ histogram. Finally, we solved the intersection of the two curves and set it as the radius. This approach meant that the selection of $\epsilon$ here was based on the density difference between phase space and possible cluster members.	

As a result, a data point with at least $k$ neighboring points was marked as the core point, and the neighbors as member points. In this step, we used three nodes in a cloud computing device to start the data running in each data set, and each node had 64 cores. For subset 1, we only used results with $\varpi$ > 2~mas, and for subset 2 we adopted all results with 1 < $\varpi$ < 2~mas. After a visual inspection, we included some results from 0.8 to 1~mas if the parallax cut did not affect the completeness of member stars.

After that, we ranked the number of member stars of each group (here, we call the initial clustering result a “group”, to avoid confusion with a star cluster) in descending order. The clustering results within the 3$\sigma$ (dispersion) range of the positions and proper motions of each group were regarded as belonging to the same group. Meanwhile, the results within a linear scale of 10 pc and a linear velocity of 1~\kms were also classified as belonging to the same group. This was used to avoid misidentifying a substructure of a known OC as a new cluster.

At last, we checked the remaining groups through visual inspection, and we removed some outliers with obvious large dispersions in their astrometric parameters. For a group with multiple dense cores, we further used the k-means algorithm~\citep{CG18} to artificially set the core number parameters and separate them into different groups.

Through the above steps, the remaining groups were taken as star cluster candidates. Similar to ~\citet{Castro18,Castro20}, we performed a statistical analysis of the results using different $k$ values. Here, the range of $k$ values was $k \in [5, 15]$. After that, we removed clusters containing fewer than 15 members. 
The frequency ($N_{Re}$) with which a given star respond in the clustering procedure under different $k$ values was also obtained. For most member stars, $N_{Re}$ could be used to measure whether they were in the densest part of a data set. Some member stars with low $N_{Re}$ values were usually at the outer edge of the data set. However, we still noticed that the photometry of many members with relatively low $N_{Re}$ values placed them on the main sequence of a color-magnitude diagram (CMD) of the corresponding cluster. 

\subsection{Isochrone fitting}\label{sec:isochrone}  

In order to determine the cluster parameters, including cluster age and extinction values, theoretical isochrones with solar metallicity~\footnote{\url{https://www.cosmos.esa.int/web/gaia/edr3-passbands}} \footnote{\url{http://stev.oapd.inaf.it/cgi-bin/cmd_3.6}} were fit to the CMD of each cluster~\citep{Bressan12}. 
The fitting function and corresponding $python$ codes were taken from the method we established in H21 and H22:
\begin{equation}\label{eqage}
\mathbf{x}_k = (G + m - M, BP-RP)_{k}
\end{equation} 

\begin{equation}\label{eqage}
\mathbf{x}_{kN} = [G_0 + c_G \cdot A_0, (BP-RP)_0 + (c_{BP_0}-c_{RP_0}) \cdot A_0]_{kN}
\end{equation}

\begin{equation}\label{eqage}
\bar{d^2}= \frac{\sum_{k=1}^{n}(\mathbf{x}_k-\mathbf{x}_{kN})^2}{n}
\end{equation}
where $\mathbf{x}_k$ is the $k_{th}$ data point in the observed photometric vector,  $\mathbf{x}_{kN}$ is the nearest point to $\mathbf{x}_k$ in the theoretical isochrone, and $\bar{d^2}$ is the average results of $n$ input $\mathbf{x}_k$ vectors. The isochrones were corrected for differential extinction and reddening values, with A$_{0}$ (at 550~nm) up to 5.00~mag. Following  ~\citet{Jordi10,Danielski18,gaia_cmd}, a polynomial function~\footnote{This product makes use of public auxiliary data provided by ESA/Gaia/DPAC/CU5 and prepared by Carine Babusiaux.} was used to compute the extinction coefficients.

Comparing the estimated distance modulus with different steps for distinct clusters in H21 and H22, the observed photometric data in this study have been corrected for distance modulus using the resolved stellar distances, which were easily derived from the reverse of the parallax values. This procedure also helped to reduce the influence caused by the different distances of member stars in the same star cluster, especially for expanding clusters (see Section~\ref{expand}).

Isochrone fittings were conducted for the high $N_{Re}$ stars, which should be less contaminated by any possible outliers. Furthermore, considering that the number of stars must be sufficient to obtain an acceptable fit, we applied the following criteria: 
\begin{itemize}
\item[\textbullet] if $n \leqslant 100$, then $N_{Re}  \geqslant 1$.
\item[\textbullet] if $n(N_{Re} \geqslant N^{*}_{Re}) \geqslant 100$, then $N_{Re}  \geqslant N^{*}_{Re}$, where $N^{*}_{Re}$ is the maximum value in the range [2,11] that satisfies the preceding conditions.
\end{itemize}

We then carried out visual inspections of each result and manually refit any isochrones that were poorly fit: we take an astrometric criteria ruwe < 1.4~\citep{Fabricius21}, and we removed the stragglers and highly reddened stars. The typical fitted accuracy was 0.05~dex in logarithmic age, and 0.05~mag in extinction. The isochrone parameters with a minimum output $\bar{d^2}$ value were adopted as the best-fitting result of a star cluster.

\section{Results and Discussion}\label{sec:results}
\subsection{Cross-matching with known clusters}\label{sec:crossmatch}

Through the above steps, we obtained a total of 886 star clusters and candidates in the whole sky. The final results contain previously known star clusters and new discoveries. The basic parameters of each cluster were taken from all member stars, including positions (median coordinates in Galactic longitude and latitude) and 1$\sigma$ dispersion of the positions; proper motions and their corresponding dispersions; and the astrophysical parameters from isochrone fits. In this part, we cross-matched the basic parameters found for our sample with the results from published works.

First, we cross-matched with the cluster catalog found by ~\citet{CG20_0} (which combined the works of ~\citet[][]{CG18,Castro18,Castro19,Castro20,Liu19}). Then we cross-matched all clusters from ~\citet[][]{Sim19,Liu19,Kounkel19,Ferreira19,Ferreira20,ferreira21,hao20,Qin20,he21,Hunt21,Casado21,li22,he22}. We also noticed a nearby cluster Group X~\citep{Oh17,Tang19} can be divided into two parts in space and kinematic, so we marked them as Group X-a and Group X-b. We adopted the same cross-match method as described in Section~\ref{clustering}, considering two cases: 

\begin{itemize}
\item[\textbullet]Known star clusters within the 3$\sigma$ dispersion ranges of the relevant astrometric parameters $(l,b,\mu_{\alpha^*},\mu_{\delta},\varpi)$ were considered to be part of the same star cluster. 
\item[\textbullet]Known star clusters within a linear scale of 10~pc and a linear velocity of 1 ~\kms were also considered as part of the same star cluster.
\end{itemize}
If one result corresponded to multiple known star clusters, we confirmed/refuted it through a visual inspection. The nearest cluster in terms of positions and kinematics was considered to be a matched cluster. 

In the pre-Gaia era, most star clusters were cataloged in the DAML02 catalog~\citep[]{Dias02} and MWSC catalog~\citep[]{Kharchenko13}, of which about 1,200 were found in Gaia DR2 by~\citet{CG18} with the UPMASK method~\citep{upmask}. Most of the remaining undiscovered clusters were thought to be in regions with large line-of-sight extinctions, or they are actually asterisms~\citep{CG20_0}. In our work, in addition to matching OCs from~\citet{CG18}, we also cross-matched the star clusters with the proper motion and distance information in these two pre-Gaia catalogs. To do this, first we extracted the results within the maximum cluster radius recorded in DAML02 and MWSC. Second, we looked for clusters within the range of three times the error in proper motion ($\sim$0.2 to 2 mas yr$^{-1}$).

\begin{figure*}
\begin{center}
	\includegraphics[width=1.\linewidth]{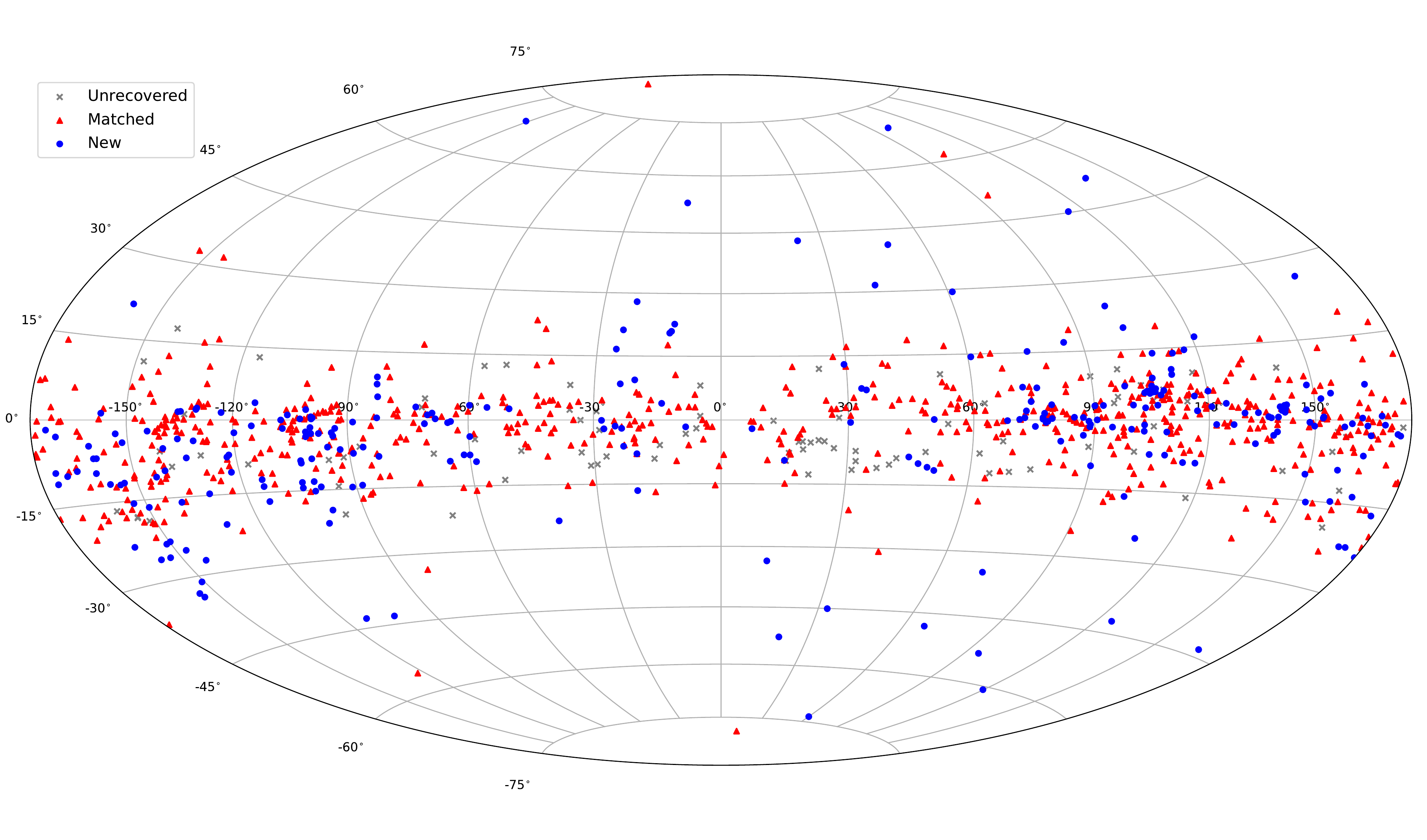}
	\caption{Locations of the new cluster candidates (blue dots) and matched known clusters (red dots) determined by the median positions of member stars in Galactic coordinates (0$^\circ$ to -180$^\circ$ represent 360$^\circ$ to 180$^\circ$ in Galactic longitude), the CG20 OCs ($\varpi$ > 0.8~mas) not recovered in this work are also marked (grey $\times$).}
	\label{fig_lb}
\end{center}
\end{figure*}

Finally, the results show that 616 clusters were matched. That is to say, our blind search in Gaia EDR3 reproduced more than 80$\%$ of the known nearby clusters. Besides, as far as we know, a total of 270 cluster candidates are not list in any published catalogs. The basic parameters of all clusters are shown in Table~1, in which column 1 gives the cluster identification of these matched known clusters and new candidates. The newly found candidates were named CWNU 1001 to CWNU 1270. Figure~\ref{fig_lb} shows the positions of matched known clusters and new clusters projected in Galactic coordinates. A full table of the parameters and the member stars for each new candidate (see examples in Table~\ref{table_mem}) are available at CDS via \url{cdsarc.u-strasbg.fr (130.79.128.5)}.

\begin{longrotatetable}
\begin{deluxetable*}{cccccccccccccc}
	\tablecaption{Derived astrophysical parameters for the 886 star clusters and candidates identified in this work. The positions, parallax and proper motions of each cluster are calculated as the median value, and the dispersion of each astrometric value are also presented. The typical fitted accuracy for log(Age/yr) and A$_0$ are 0.05 dex and 0.05}~mag, respectively.
	\label{tab_all}
	\tablewidth{1.0pt}
	\tabletypesize{\scriptsize}
\tablehead{ Cluster  & $l$ & $\sigma_{l}$ & $b$ & $\sigma_{b}$ & n & $\varpi$ & $\sigma_{\varpi}$ & $\mu_{\alpha^*}$ & $\sigma_{\mu_{\alpha^*}}$ & $\mu_{\delta}$ & $\sigma_{\mu_{\delta}}$ & log(Age/yr) & A$_0$\\ 
 & ($^{\circ}$) & ($^{\circ}$) & ($^{\circ}$) & ($^{\circ}$) &  & (mas) & (mas) & (mas~  yr$^{-1}$) & (mas~  yr$^{-1}$) & (mas~  yr$^{-1}$) & (mas~  yr$^{-1}$)  &   & mag}
	\startdata
Alessi 1 & 123.263 & 0.118 & -13.360 & 0.100 & 46 & 1.41 & 0.04 & 6.48 & 0.10 & -6.42 & 0.15 & 8.70  & 0.60\\ 
Alessi 10 & 31.626 & 0.272 & -21.041 & 0.169 & 81 & 2.26 & 0.05 & 1.45 & 0.17 & -7.90 & 0.21 & 7.90  & 0.60\\ 
Alessi 12 & 67.448 & 0.400 & -11.476 & 0.357 & 198 & 1.84 & 0.05 & 4.32 & 0.19 & -4.70 & 0.15 & 7.90  & 0.35\\ 
Alessi 13 & 237.237 & 2.393 & -55.714 & 1.100 & 66 & 9.54 & 0.30 & 36.14 & 1.60 & -4.61 & 0.98 & 7.50  & 0.05\\ 
Alessi 19 & 40.061 & 0.278 & 12.633 & 0.177 & 43 & 1.69 & 0.04 & -1.00 & 0.13 & -7.18 & 0.14 & 7.25  & 0.30\\ 
Alessi 2 & 152.333 & 0.272 & 6.367 & 0.249 & 140 & 1.62 & 0.04 & -0.94 & 0.15 & -1.09 & 0.14 & 8.45  & 0.85\\ 
Alessi 20 & 117.242 & 0.766 & -3.763 & 0.448 & 407 & 2.33 & 0.08 & 8.02 & 0.47 & -2.47 & 0.41 & 7.05  & 1.05\\ 
Alessi 21 & 223.461 & 0.286 & -0.021 & 0.239 & 122 & 1.76 & 0.05 & -5.53 & 0.19 & 2.65 & 0.18 & 7.65  & 0.40\\ 
Alessi 24 & 329.017 & 0.407 & -14.594 & 0.257 & 149 & 2.07 & 0.05 & -0.46 & 0.17 & -8.95 & 0.18 & 7.85  & 0.30\\ 
Alessi 3 & 257.554 & 0.998 & -15.247 & 0.647 & 162 & 3.61 & 0.11 & -9.82 & 0.48 & 11.96 & 0.51 & 8.60  & 0.50\\ 
Alessi 31 & 15.355 & 0.137 & 7.670 & 0.106 & 69 & 1.48 & 0.03 & -1.14 & 0.15 & -4.21 & 0.13 & 7.60  & 2.70\\ 
Alessi 37 & 101.701 & 0.211 & -11.403 & 0.168 & 98 & 1.41 & 0.03 & 0.35 & 0.09 & -1.68 & 0.11 & 7.90 & 0.88\\ 
Alessi 43 & 262.531 & 0.319 & 1.492 & 0.166 & 404 & 1.06 & 0.05 & -5.57 & 0.39 & 3.92 & 0.38 & 6.55  & 1.15\\ 
Alessi 5 & 287.982 & 0.663 & -1.947 & 0.405 & 491 & 2.50 & 0.08 & -15.45 & 0.41 & 2.56 & 0.49 & 7.60  & 0.50\\ 
Alessi 6 & 313.655 & 0.116 & -5.545 & 0.091 & 119 & 1.13 & 0.03 & -10.56 & 0.14 & -5.53 & 0.14 & 8.55  & 1.10\\ 
... & ... & ... & ... & ... & ... & ... & ... & ... & ... & ... & ... & ... & ...\\
CWNU 1001 & 139.019 & 0.832 & -2.281 & 0.278 & 38 & 2.64 & 0.10 & 13.55 & 1.04 & -15.24 & 1.10 & 8.65  & 0.90\\ 
CWNU 1002 & 104.656 & 0.129 & 5.760 & 0.064 & 34 & 0.94 & 0.07 & -2.83 & 0.67 & -2.70 & 0.94 & 7.70  & 1.25\\ 
CWNU 1003 & 206.572 & 0.165 & -2.546 & 0.272 & 37 & 1.54 & 0.03 & -1.77 & 0.16 & -4.76 & 0.11 & 7.55  & 0.40\\ 
CWNU 1004 & 336.145 & 0.363 & 8.447 & 0.316 & 24 & 6.26 & 0.06 & -11.10 & 0.36 & -23.52 & 0.45 & 7.05  & 2.95\\ 
CWNU 1005 & 77.763 & 0.107 & -0.527 & 0.052 & 20 & 0.88 & 0.02 & 0.09 & 0.96 & -3.55 & 1.27 & 7.65  & 2.35\\ 
CWNU 1006 & 120.186 & 0.680 & 5.893 & 0.276 & 26 & 3.00 & 0.04 & -6.62 & 0.26 & 0.14 & 0.27 & 7.85 & 0.75\\ 
CWNU 1007 & 185.317 & 0.758 & -11.062 & 0.543 & 87 & 2.87 & 0.30 & -0.10 & 0.46 & -5.89 & 0.47 & 7.55  & 1.25\\ 
CWNU 1008 & 291.971 & 0.334 & 0.264 & 0.256 & 28 & 2.16 & 0.05 & -14.32 & 0.45 & -2.85 & 0.27 & 7.50  & 0.65\\
... & ... & ... & ... & ... & ... & ... & ... & ... & ... & ... & ... & ... & ...
	\enddata
\end{deluxetable*}
\end{longrotatetable}
%

\begin{longrotatetable}
\begin{deluxetable*}{cccccccccccccc}
	\tablecaption{The astrometric and photometric parameters of the member stars from Gaia EDR3~\citep{gaia2021}. The N$_{Re}$(from 1 to 11) is the number of responses derived from DBSCAN (see Section~\ref{clustering}).}
	\tablewidth{0pt}
	\label{table_mem}
	\tabletypesize{\scriptsize}
\tablehead{ source$\_$id&	l	&b	&parallax&	parallax$\_$error	&pmra	&pmra$\_$error&	pmdec&	pmdec$\_$error& ruwe	&phot$\_$g$\_$mean$\_$mag	&bp$\_$rp  & N$_{Re}$&Cluster\\ 
   & ($^{\circ}$)  & ($^{\circ}$)  & (mas) & (mas) & (mas~yr$^{-1}$) & (mas~yr$^{-1}$) & (mas~yr$^{-1}$) & (mas~yr$^{-1}$)  & &(mag)&(mag)&  & }
	\startdata
414508878584171904 & 123.115 & -13.420 & 1.42 & 0.01 & 6.48 & 0.01 & -6.40 & 0.01 & 1.04 & 12.61 & 0.65 & 11 & Alessi 1\\ 
414511524284025728 & 123.038 & -13.430 & 1.50 & 0.07 & 6.56 & 0.06 & -6.44 & 0.05 & 1.07 & 16.70 & 1.67 & 8 & Alessi 1\\ 
414514646721724032 & 123.043 & -13.248 & 1.33 & 0.04 & 6.35 & 0.04 & -6.44 & 0.04 & 3.12 & 11.95 & 0.62 & 4 & Alessi 1\\ 
414509119102333184 & 123.064 & -13.402 & 1.44 & 0.02 & 6.48 & 0.02 & -6.46 & 0.02 & 1.29 & 12.02 & 0.55 & 11 & Alessi 1\\ 
414509084742597248 & 123.066 & -13.409 & 1.45 & 0.02 & 6.62 & 0.02 & -6.53 & 0.02 & 1.02 & 11.46 & 0.54 & 11 & Alessi 1\\ 
414509084742596992 & 123.068 & -13.408 & 1.33 & 0.02 & 6.45 & 0.02 & -6.35 & 0.02 & 0.96 & 14.88 & 1.23 & 6 & Alessi 1\\ 
402508190203269376 & 123.268 & -13.298 & 1.41 & 0.03 & 6.36 & 0.03 & -6.43 & 0.02 & 1.30 & 10.51 & 0.35 & 11 & Alessi 1\\ 
414510660992079232 & 123.091 & -13.380 & 1.45 & 0.02 & 6.41 & 0.01 & -6.29 & 0.01 & 1.04 & 12.65 & 0.66 & 11 & Alessi 1\\ 
414510459132125824 & 123.105 & -13.380 & 1.38 & 0.03 & 6.45 & 0.03 & -6.42 & 0.02 & 0.99 & 15.25 & 1.14 & 11 & Alessi 1\\ 
414509978093147136 & 123.190 & -13.364 & 1.44 & 0.03 & 6.55 & 0.03 & -6.42 & 0.02 & 1.69 & 10.81 & 0.44 & 11 & Alessi 1\\ 
414518877267971456 & 123.225 & -13.165 & 1.39 & 0.02 & 6.45 & 0.01 & -6.31 & 0.02 & 1.46 & 13.24 & 0.79 & 11 & Alessi 1\\ 
402505991180022528 & 123.249 & -13.390 & 1.47 & 0.02 & 6.48 & 0.02 & -6.37 & 0.02 & 1.52 & 9.82 & 1.17 & 11 & Alessi 1\\ 
414516506447331200 & 123.195 & -13.290 & 1.42 & 0.05 & 6.24 & 0.04 & -6.66 & 0.03 & 1.58 & 10.88 & 0.42 & 11 & Alessi 1\\ 
... & ... & ... & ... & ... & ... & ... & ... & ... & ... & ... & ... & ... & ...\\ 
4191234391174022016 & 32.068 & -21.628 & 2.24 & 0.10 & 1.87 & 0.10 & -7.63 & 0.07 & 1.04 & 17.45 & 2.55 & 10 & Alessi 10\\ 
4190657903483144960 & 30.661 & -21.229 & 2.24 & 0.11 & 0.89 & 0.10 & -8.15 & 0.07 & 1.07 & 17.33 & 2.61 & 5 & Alessi 10\\ 
4191081451680369152 & 31.933 & -20.778 & 2.17 & 0.12 & 1.39 & 0.11 & -7.81 & 0.08 & 0.96 & 17.61 & 2.46 & 11 & Alessi 10\\ 
4191074549670320896 & 31.769 & -20.852 & 2.29 & 0.07 & 1.31 & 0.06 & -7.72 & 0.04 & 1.32 & 15.59 & 1.62 & 11 & Alessi 10\\ 
4191070529580877568 & 31.979 & -21.029 & 2.18 & 0.08 & 1.83 & 0.08 & -7.64 & 0.06 & 0.95 & 17.02 & 2.19 & 11 & Alessi 10\\ 
4190669036038417152 & 30.982 & -21.322 & 2.17 & 0.03 & 1.12 & 0.03 & -8.65 & 0.02 & 0.99 & 14.79 & 1.40 & 11 & Alessi 10\\ 
4190818874560007552 & 31.318 & -21.230 & 2.12 & 0.10 & 1.43 & 0.11 & -8.10 & 0.07 & 0.94 & 17.55 & 2.71 & 11 & Alessi 10\\ 
4190723255705770240 & 30.784 & -20.945 & 2.24 & 0.06 & 1.31 & 0.05 & -7.17 & 0.04 & 3.44 & 12.46 & 0.86 & 11 & Alessi 10\\ 
4191033524142530688 & 31.642 & -21.036 & 2.29 & 0.07 & 1.14 & 0.06 & -8.33 & 0.05 & 1.24 & 15.98 & 2.31 & 11 & Alessi 10\\ 
4190732120518267776 & 31.029 & -21.061 & 2.33 & 0.07 & 1.45 & 0.07 & -8.19 & 0.04 & 0.98 & 16.39 & 1.94 & 11 & Alessi 10\\ 
4190738820667368960 & 31.156 & -20.957 & 2.28 & 0.03 & 1.48 & 0.03 & -7.96 & 0.02 & 0.98 & 14.50 & 1.24 & 11 & Alessi 10\\  
... & ... & ... & ... & ... & ... & ... & ... & ... & ... & ... & ... & ... & ...\\ 
460642801638611840 & 139.488 & -1.900 & 2.69 & 0.11 & 11.82 & 0.10 & -15.57 & 0.12 & 1.06 & 17.57 & 2.80 & 5 & CWNU 1001\\ 
454786252955059968 & 138.176 & -2.661 & 2.62 & 0.01 & 13.49 & 0.01 & -15.32 & 0.02 & 1.02 & 13.70 & 1.16 & 2 & CWNU 1001\\ 
459869986704520704 & 139.824 & -2.105 & 2.45 & 0.02 & 12.22 & 0.02 & -13.60 & 0.02 & 0.98 & 13.99 & 1.39 & 4 & CWNU 1001\\ 
460870885883240704 & 137.697 & -2.062 & 2.62 & 0.04 & 13.84 & 0.04 & -14.67 & 0.04 & 0.98 & 16.03 & 1.91 & 4 & CWNU 1001\\ 
454845218566439808 & 137.832 & -2.456 & 2.71 & 0.01 & 14.40 & 0.01 & -15.44 & 0.02 & 0.95 & 12.86 & 1.02 & 4 & CWNU 1001\\ 
460605388184245376 & 138.883 & -2.211 & 2.83 & 0.02 & 13.61 & 0.02 & -16.57 & 0.03 & 1.04 & 14.65 & 1.47 & 5 & CWNU 1001\\ 
460799829945182976 & 138.115 & -2.358 & 2.60 & 0.02 & 12.56 & 0.01 & -16.90 & 0.02 & 0.89 & 11.33 & 0.84 & 3 & CWNU 1001\\ 
... & ... & ... & ... & ... & ... & ... & ... & ... & ... & ... & ... & ... & ...

	\enddata
\end{deluxetable*}
\end{longrotatetable}

\subsection{Matched clusters}\label{expand}
\begin{figure*}
\begin{center}
	\includegraphics[width=0.9\linewidth]{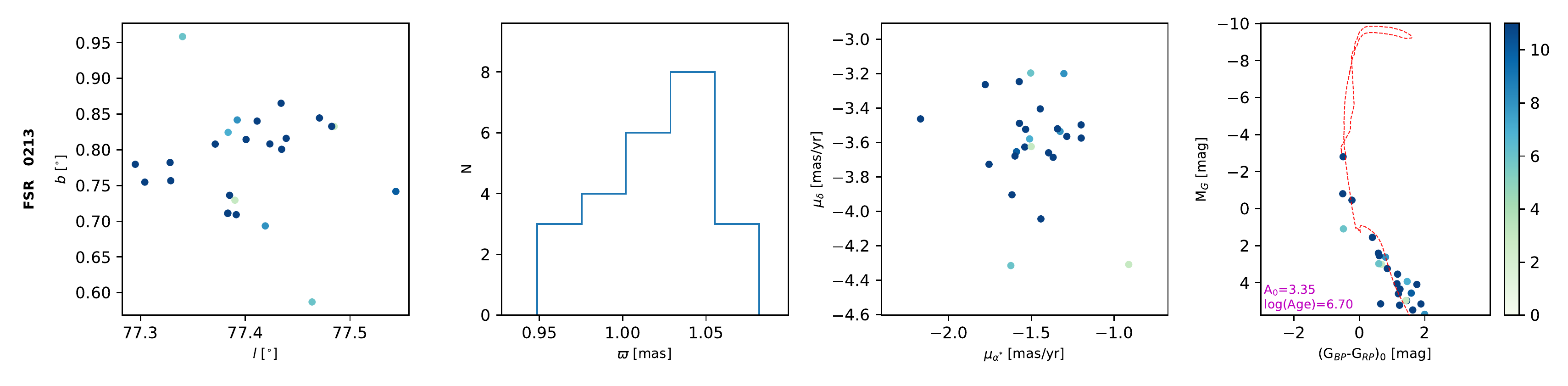}
	\includegraphics[width=0.9\linewidth]{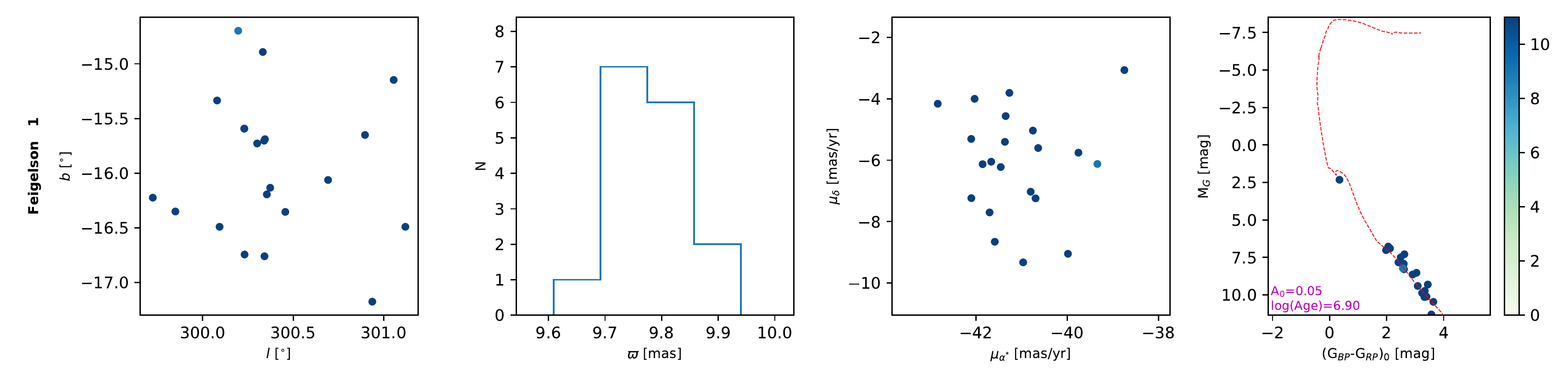}
	\caption{Astrometric and photometric examples of the newly identified pre-Gaia OCs. Left to right: spatial distribution, parallax histogram, proper motion distribution, CMD with best-fitting isochrone line. The color bars represent the number of responses in the clustering, $N_{Re}$, of each cluster member.}
	\label{fig_pregaia}
\end{center}
\end{figure*}

In total, the 616 matched clusters contain about 123,408 stars that responded at least once in DBSCAN. Figure~\ref{fig_pregaia} shows two matched clusters, FSR 0213 and Feigelson 1. The former is a star cluster found by 2MASS near-infrared observations~\citep[][]{Skrutskie06,Froebrich07}, which has a relatively high extinction; the latter is a young OC located very close to the Solar System in the southern sky~\citep[]{Dias02}, which has a large proper motion and was absent from previous Gaia star cluster searches. In this study, 10 pre-Gaia OCs were newly identified. 

%
%

\begin{figure*}
\begin{center}
	\includegraphics[width=0.9\linewidth]{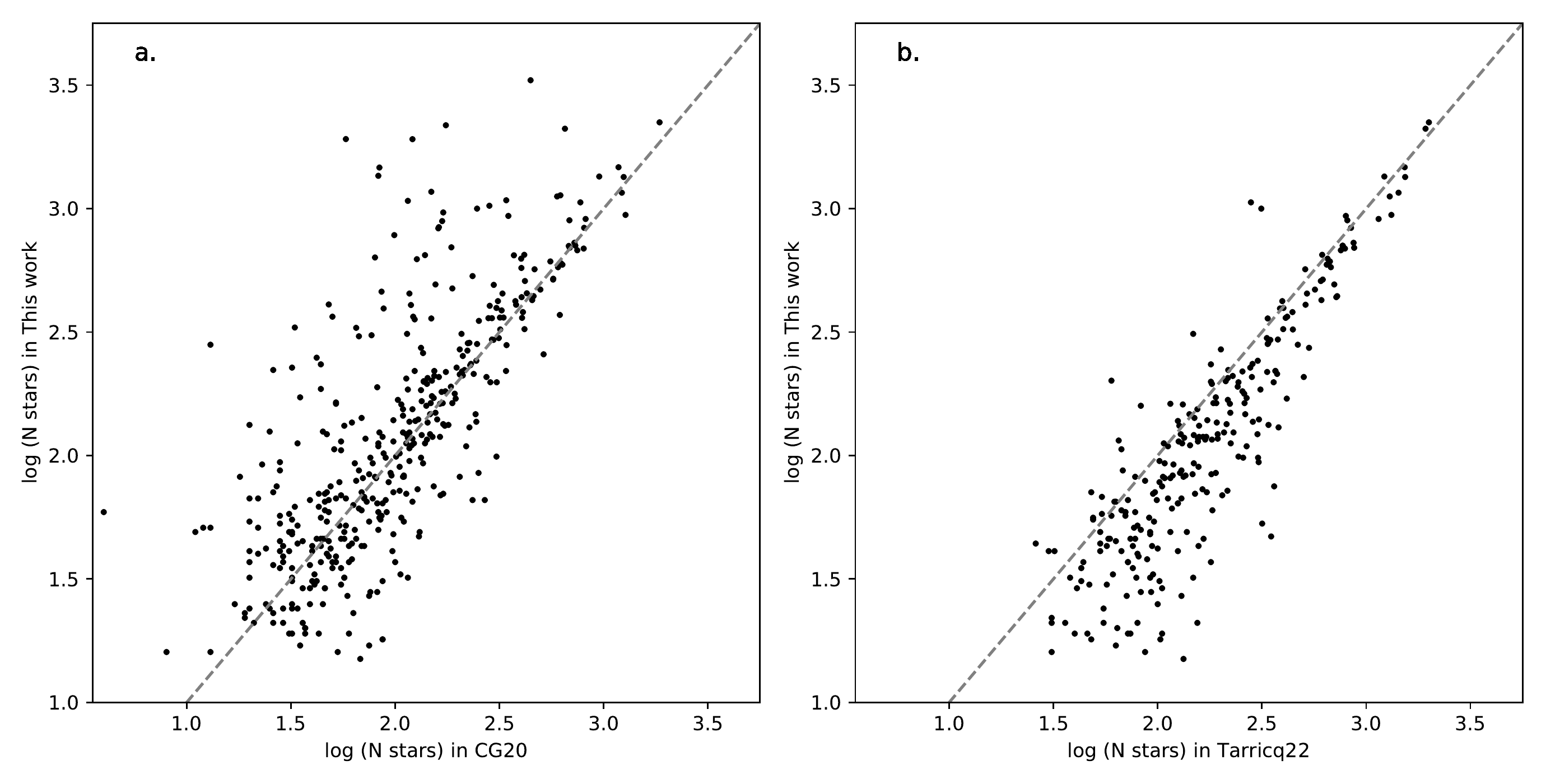}
	\caption{Difference between the number of member stars determined in this study and those found by CG20 and Tarricq22. The dashed line indicates that the two axes have equal values.}
	\label{fig_nstar}
\end{center}
\end{figure*}

\begin{figure*}
\begin{center}
	\includegraphics[width=0.9\linewidth]{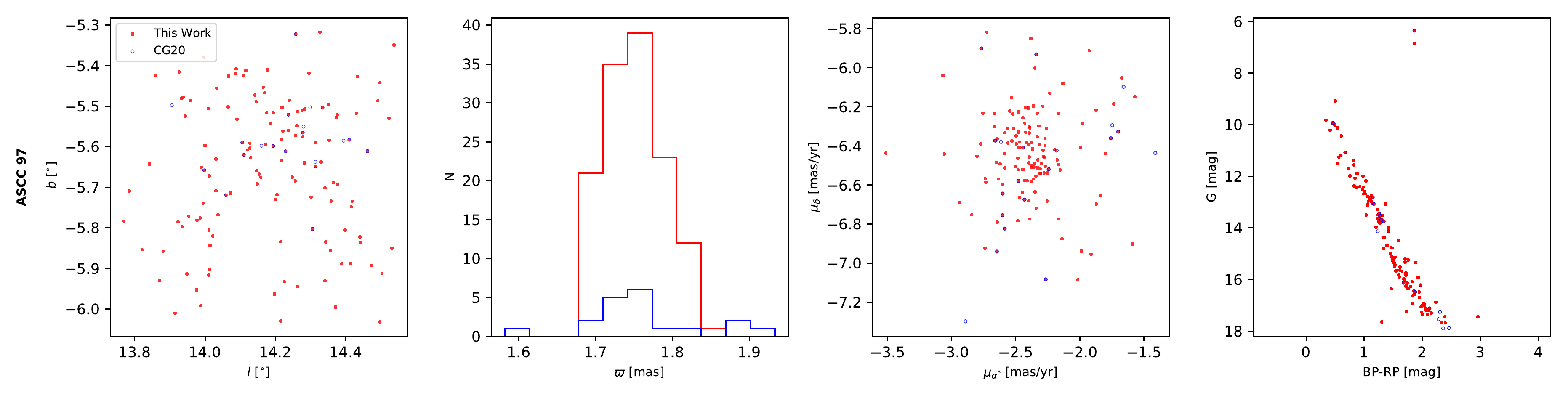}	
	\includegraphics[width=0.9\linewidth]{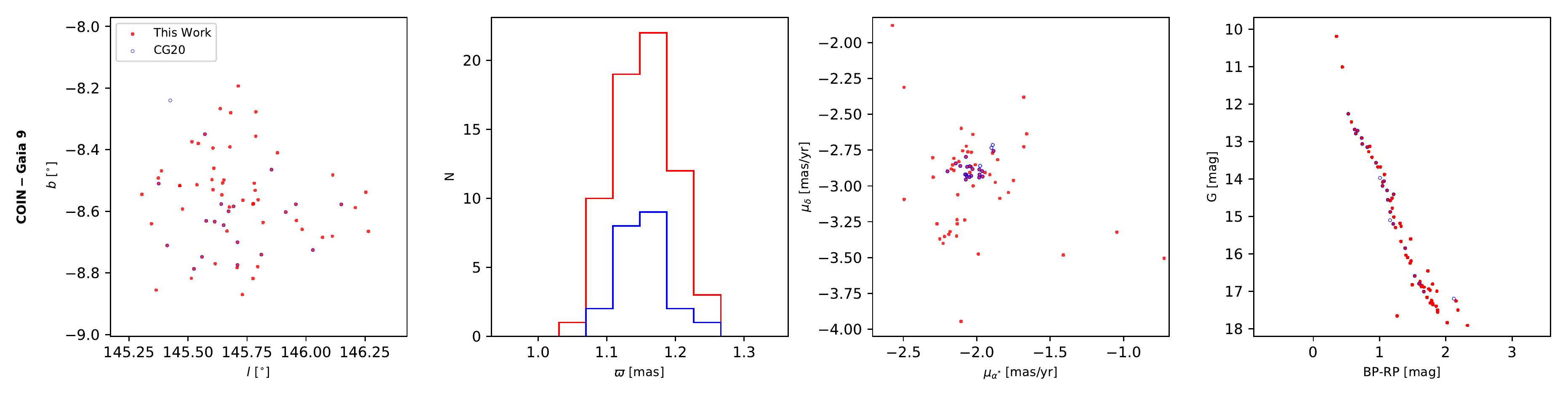}
\includegraphics[width=0.9\linewidth]{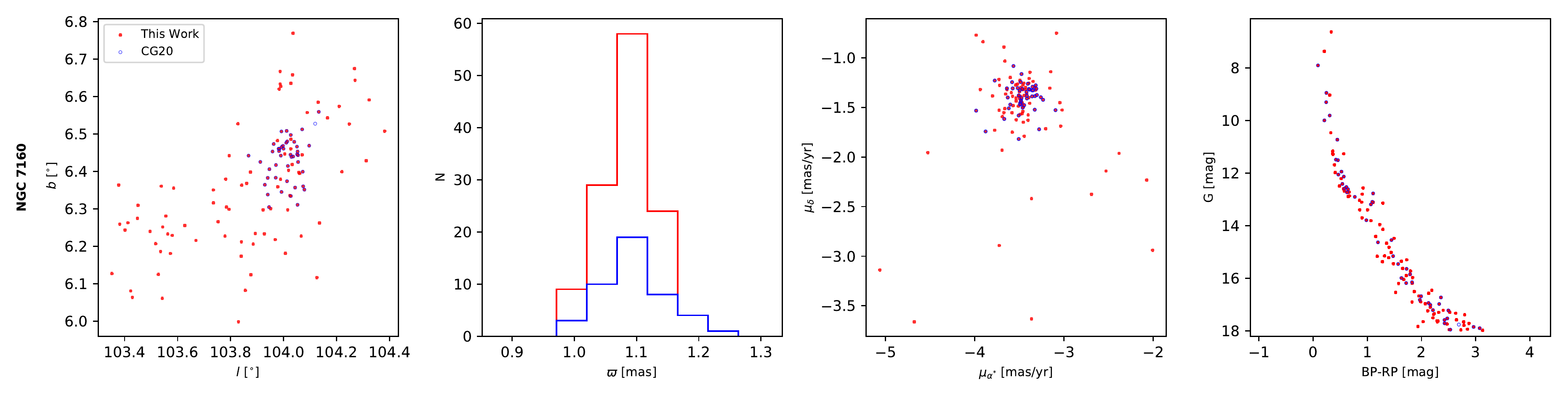}
	\caption{Examples of comparison of our results (in red) with CG20 (in blue). The four subplots show the spatial distribution, parallax statistics, proper motion distribution, and CMD of the cluster members.}
	\label{fig_cg}
\end{center}
\end{figure*}

For most of the matched clusters, we have found more member stars than clusters in Gaia DR2 (Fig.~\ref{fig_nstar}a). As a comparison, for the matched 408 OCs in CG20 ($\varpi \geqslant$ 0.8 mas), the total number of newly detected members is 99,524. This corresponded to about 243 stars per cluster, representing an increase of about 50$\%$ membership relative to CG20 (in total, 66,803 stars, and 163 stars per cluster). Figure~\ref{fig_cg} presents three examples of our new products, including NGC~7160, ASCC~97, and a recently discovered OC COIN-Gaia~9. For all examples the parallax ranges of the cluster members have not changed significantly, but our work has found more member stars. For ASCC~97, our results have extend its spatial structure, for COIN-Gaia~9, we extend its kinematic structure, while for NGC~7160, both its spatial and kinematic structures are extended.


\begin{figure*}
\begin{center}
   \includegraphics[width=0.9\linewidth]{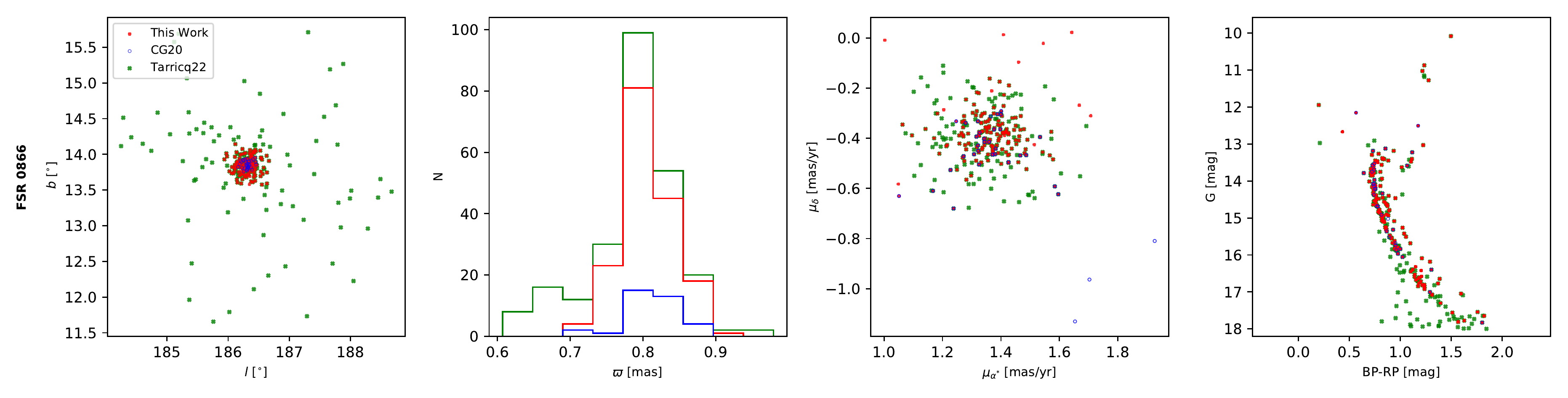}
    \includegraphics[width=0.9\linewidth]{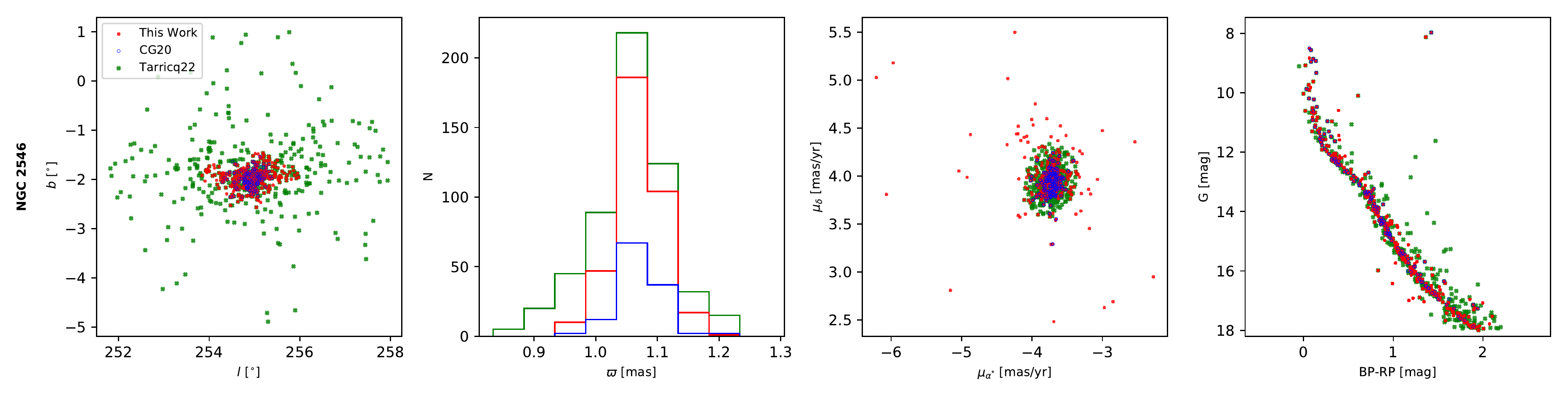}
	\includegraphics[width=0.9\linewidth]{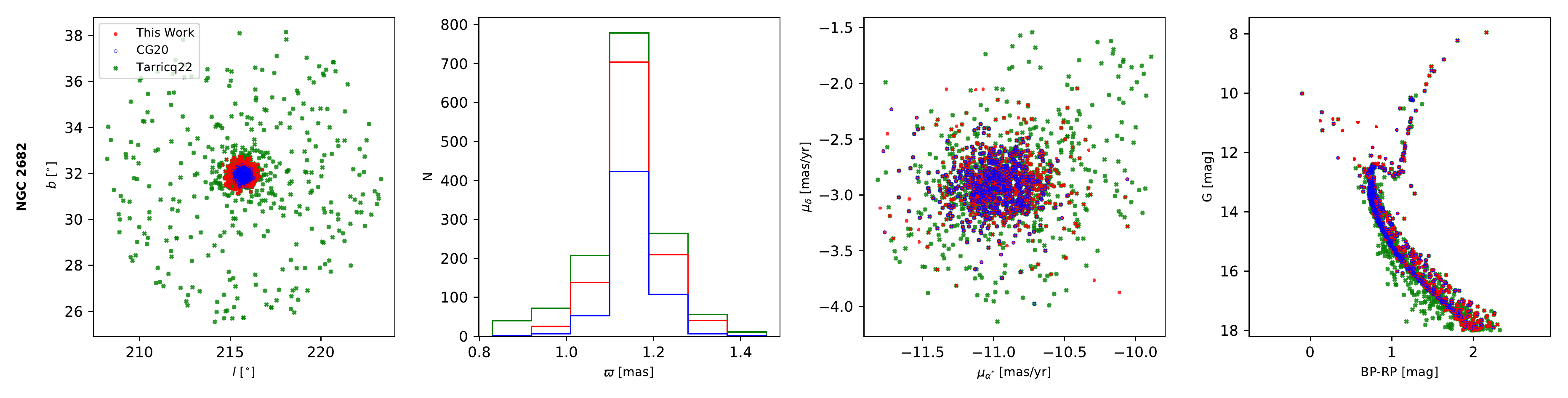}
	\caption{Same as Fig.~\ref{fig_cg}, but for the examples of comparison of our results (in red) with CG20 (in blue), and Tarricq22 (in green).}	
	\label{fig_tar}
\end{center}
\end{figure*}

A recent study of cluster membership~\citep[][hereafter Tarricq22]{Tarricq22} also showed that member stars of many nearby clusters in CG20 were underestimated. 
In our results, 257 star clusters are cross matched with Tarricq22. As shown in Fig.~\ref{fig_nstar}b. In our results, most of the cluster members have fewer stars than Tarricq22, where the total number of the member stars is 54,212 and 67,262, respectively, for ours and their study.
Figure~\ref{fig_tar} shows three well known OCs, including NGC~2682, NGC~2456, and FSR~0866. In these examples, compared with CG20, both our results and those of Tarricq22 extend the cluster halos. Our results mostly coincide with the central part of the cluster members in Tarricq22 , and the latter extends more on the parallax and space of cluster member stars. However, it can be seen that the CMDs in our results are cleaner. 
This is probably because most of member stars in our work are located in more concentrated areas and are therefore less contaminated by field stars.
In addition, there are also differences in some details, for NGC~2546, the kinematics of members in our work is more extended than in Tarricq22; and more blue/yellow stragglers in NGC~2682~\citep{Deng99,Geller15} can be seen in our work.

\begin{figure*}
\begin{center}
	\includegraphics[width=0.45\linewidth]{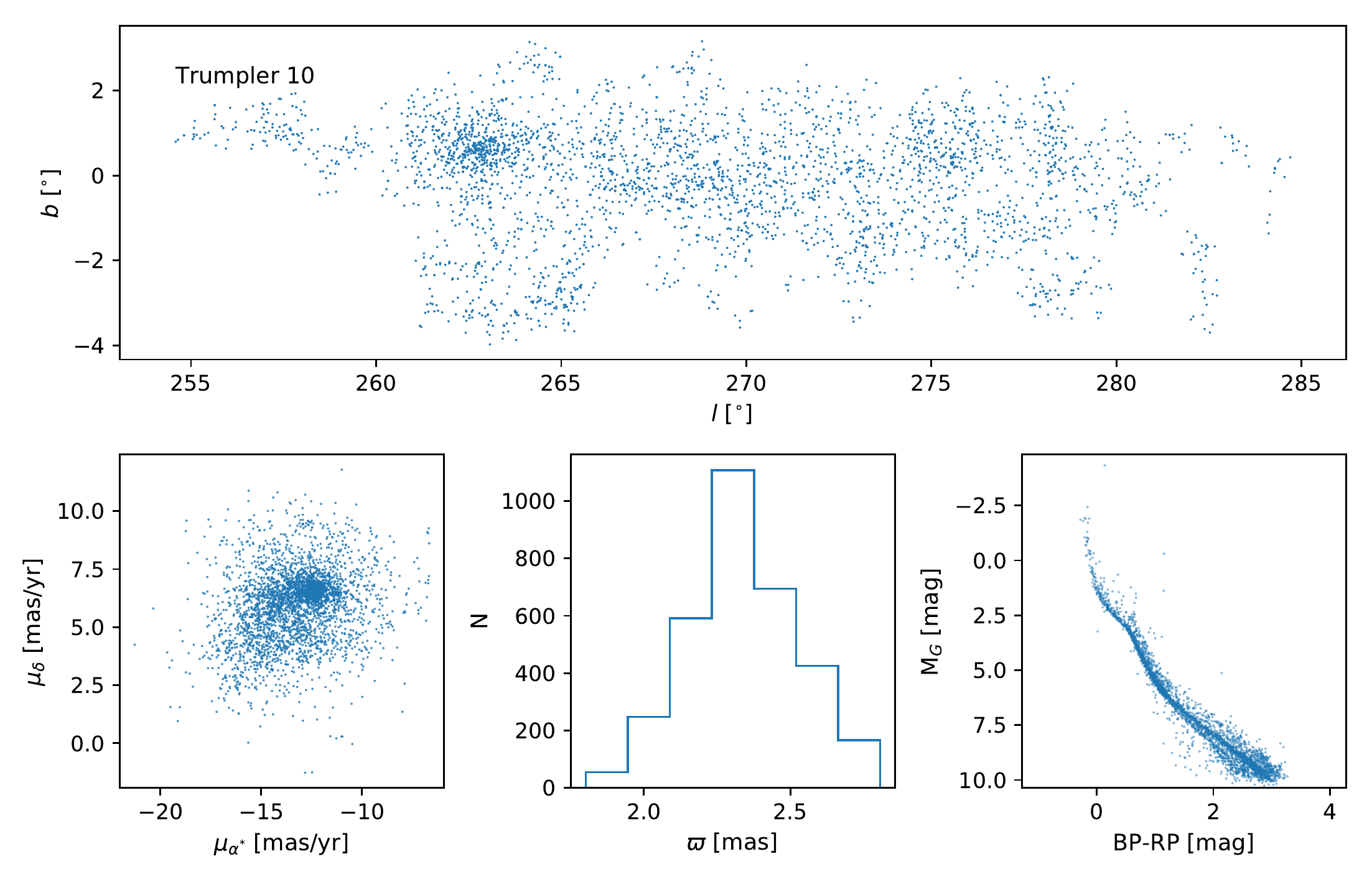}
	\includegraphics[width=0.45\linewidth]{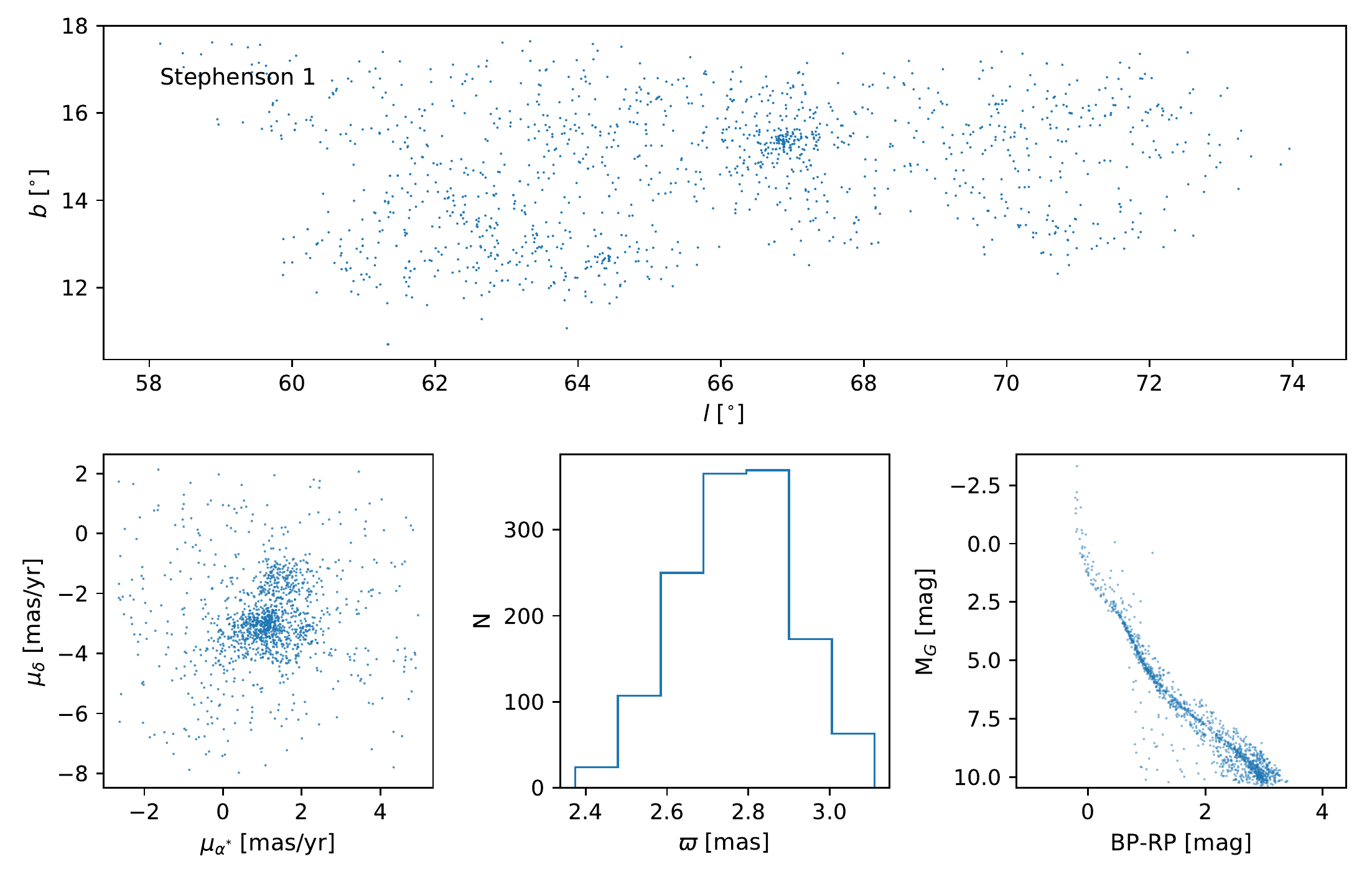}
	\includegraphics[width=0.45\linewidth]{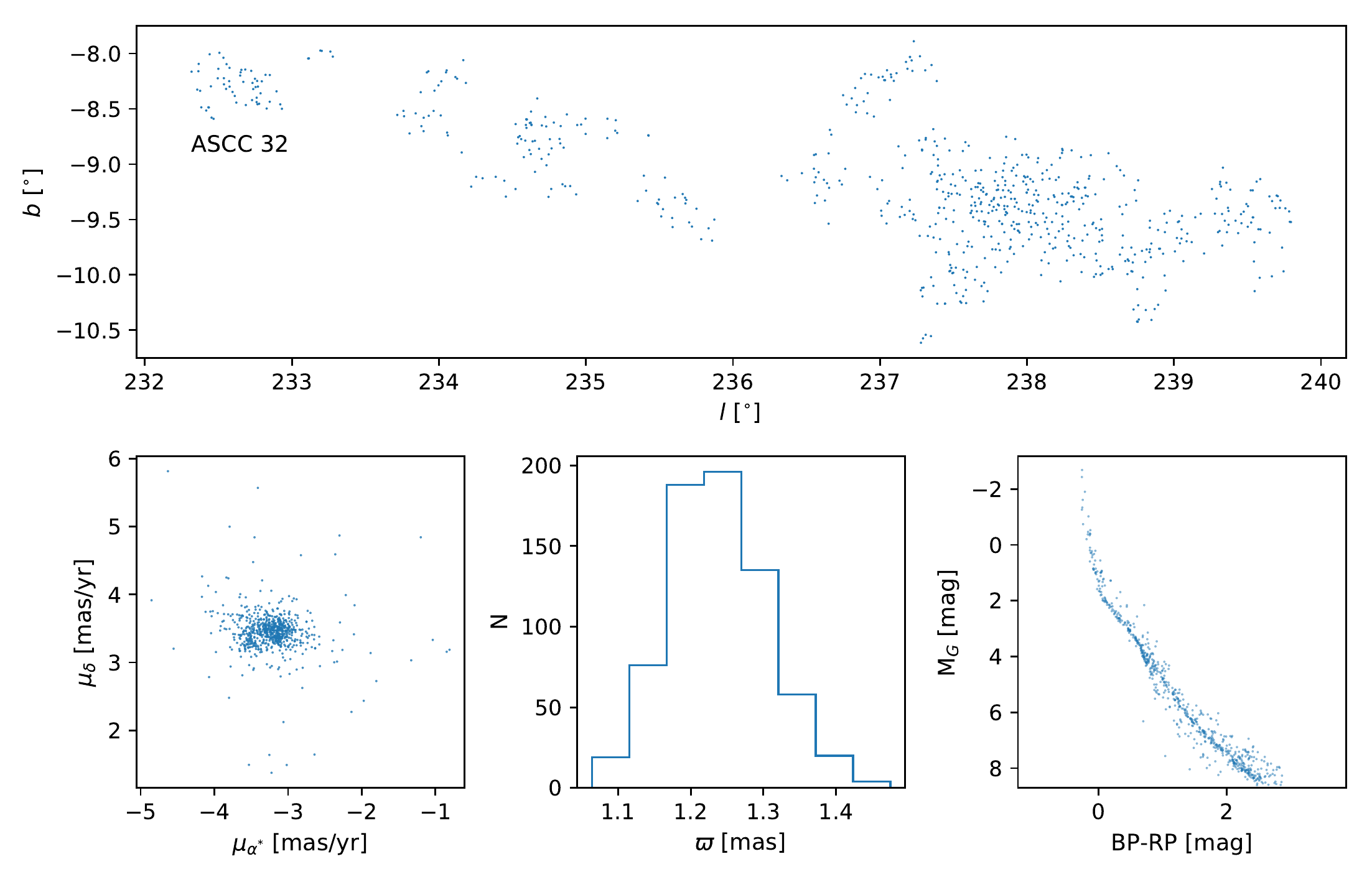}
	\includegraphics[width=0.45\linewidth]{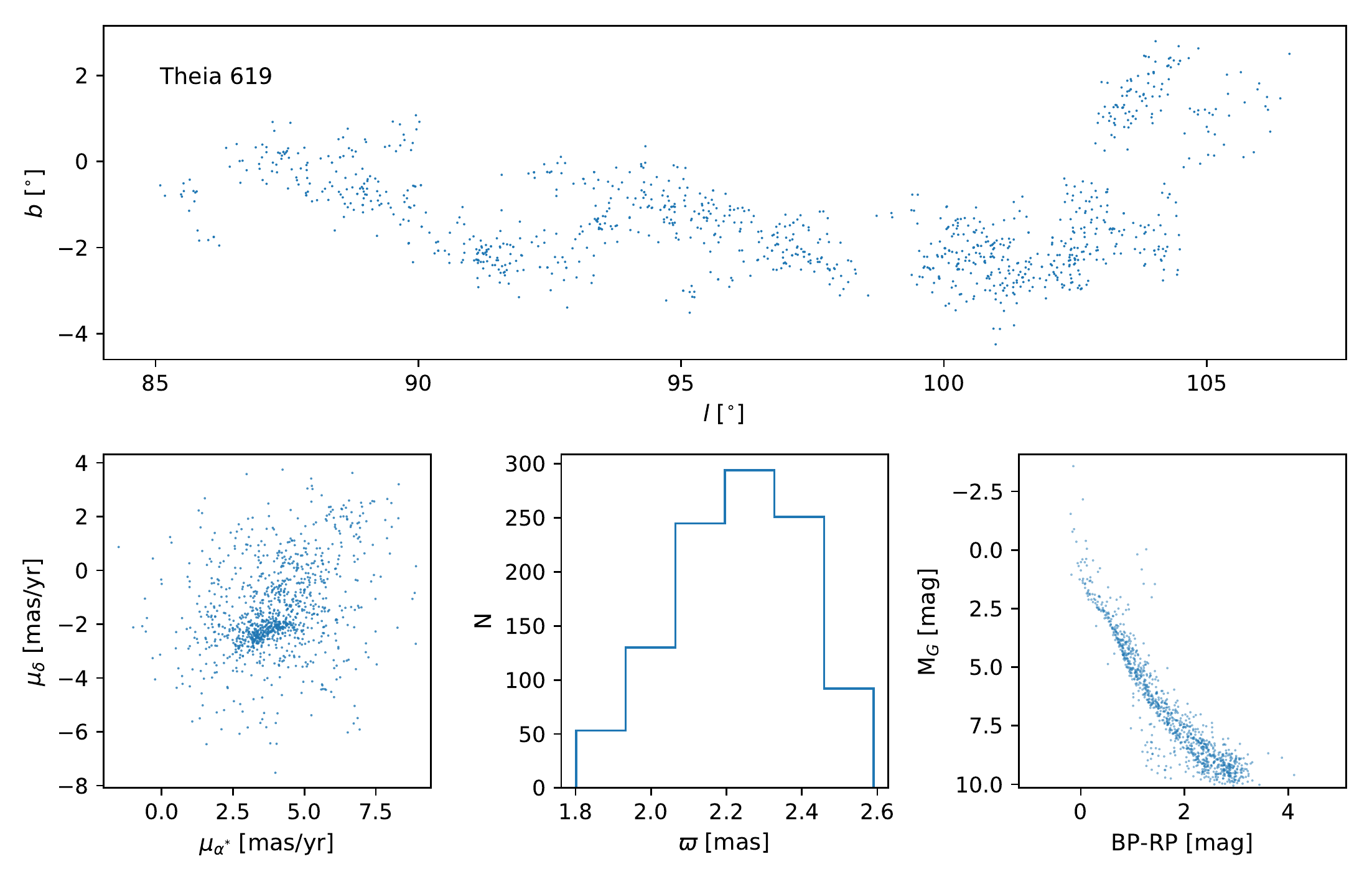}
	\caption{Astrometric and photometric examples for the matched expanding clusters. The upper panel in each subplot} presents the spatial distribution of the cluster members, the lower three panels show the parallax statistics, proper motion distribution, and CMD of the cluster.
	\label{fig_trump}
\end{center}
\end{figure*}

At the same time, the expanding structures of some clusters were also detected. Figure~\ref{fig_trump} presents four examples for Trumpler~10, Stephenson~1, ASCC~32, and Theia~619. Trumpler~10 and Stephenson~1 were identified as an expanding cluster or sparse association in Gaia DR2~\citep[][]{Cantat19_exp, Kounkel19}, where both of them have a visible core. In contrast, ASCC~32 and Theia~619 have no obvious cores, but they possess elongated filamentous extension structures, which may spread up to a hundred parsecs. Due to their close distances and being less affected by extinction, member stars with sparser distributions were revealed more richly through the clustering method. However, ASCC~32 and Theia~619 have discontinuities in space, which may be caused by their low stellar density. It can also be seen that their CMDs presented young cluster main sequences, which showed that the members in each cluster originated from the same molecular cloud tens of millions of years ago.

 \subsection{ New cluster candidates}
 
 \subsubsection{ Classifications}\label{sec:classification}
 
\begin{figure*}
\begin{center}
	\includegraphics[width=.9\linewidth]{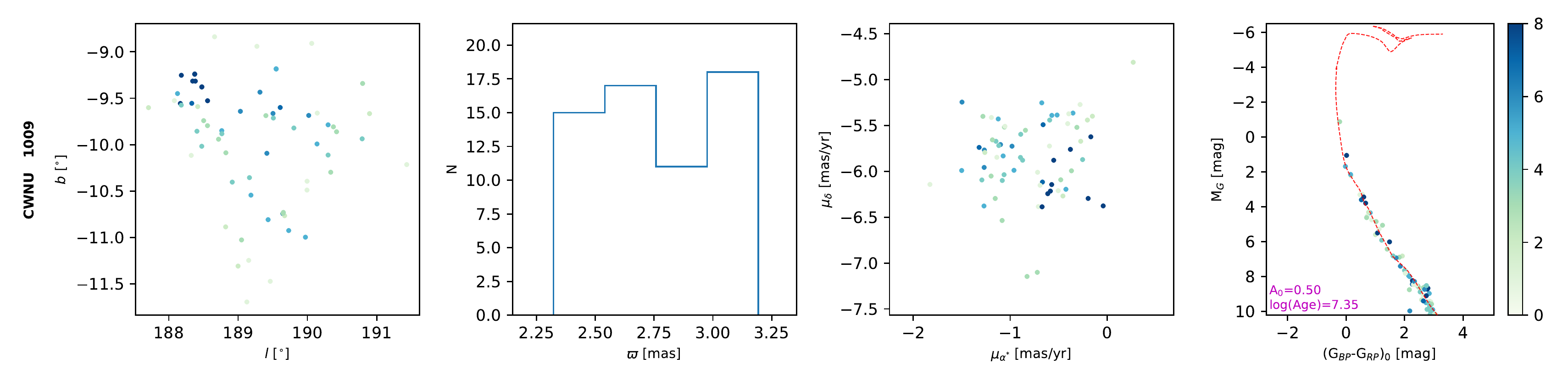}
	\includegraphics[width=.9\linewidth]{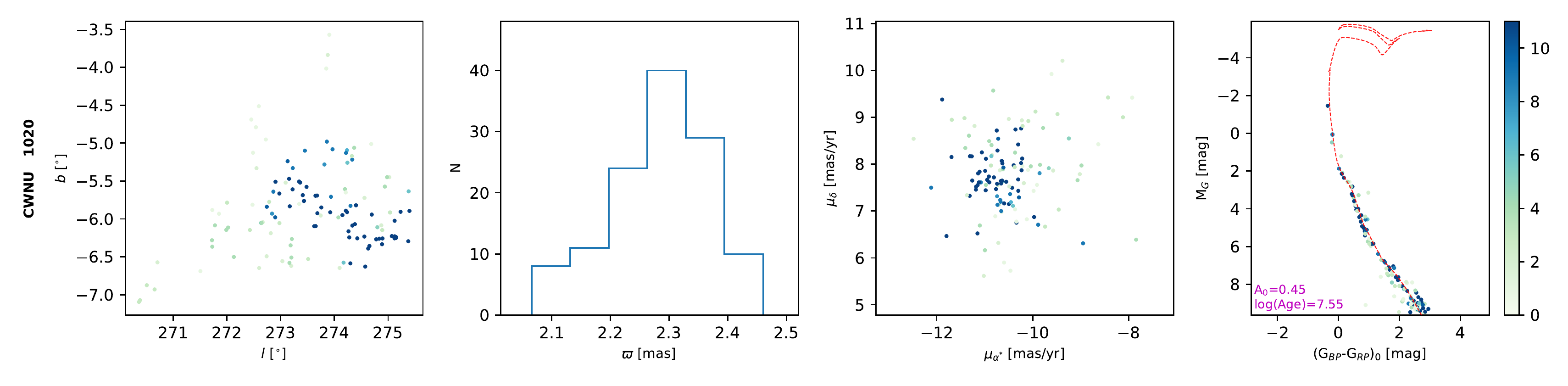}
	\includegraphics[width=.9\linewidth]{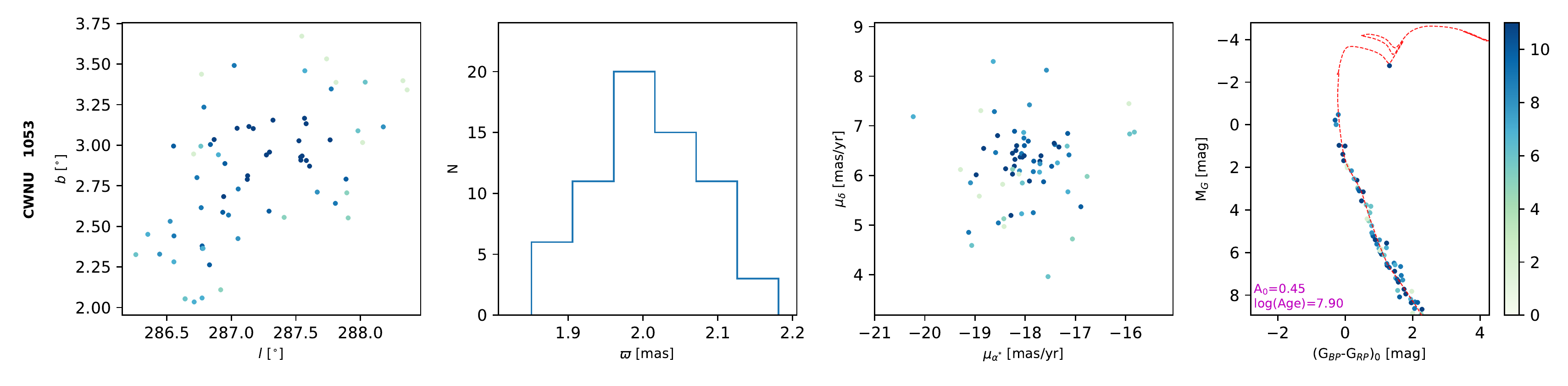}
	\includegraphics[width=.9\linewidth]{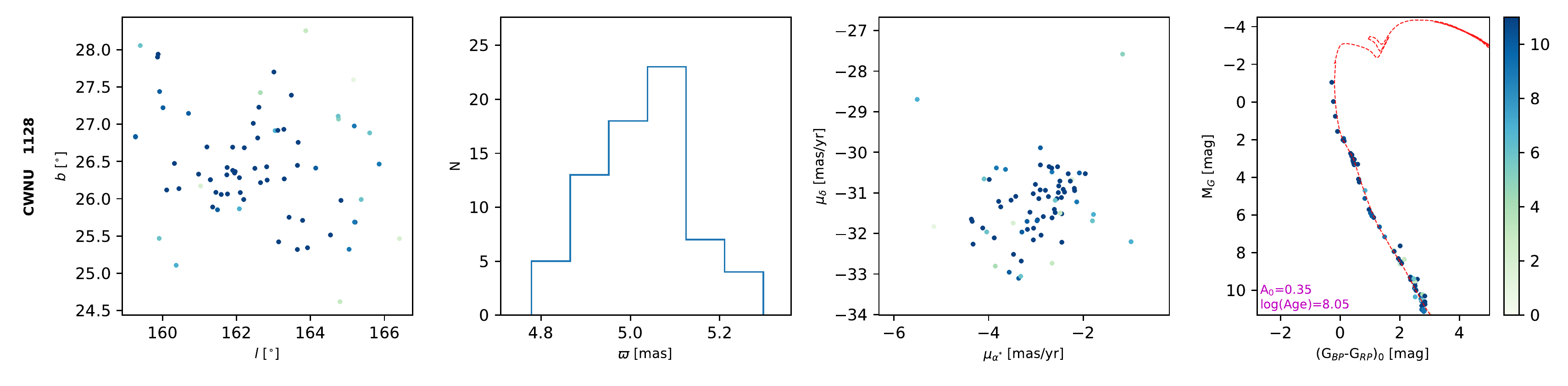}
	\caption{Same as Fig.~\ref{fig_pregaia}, but for the new candidates in class 1.}
	\label{fig_class1}
\end{center}
\end{figure*}

\begin{figure*}
\begin{center}
	\includegraphics[width=0.9\linewidth]{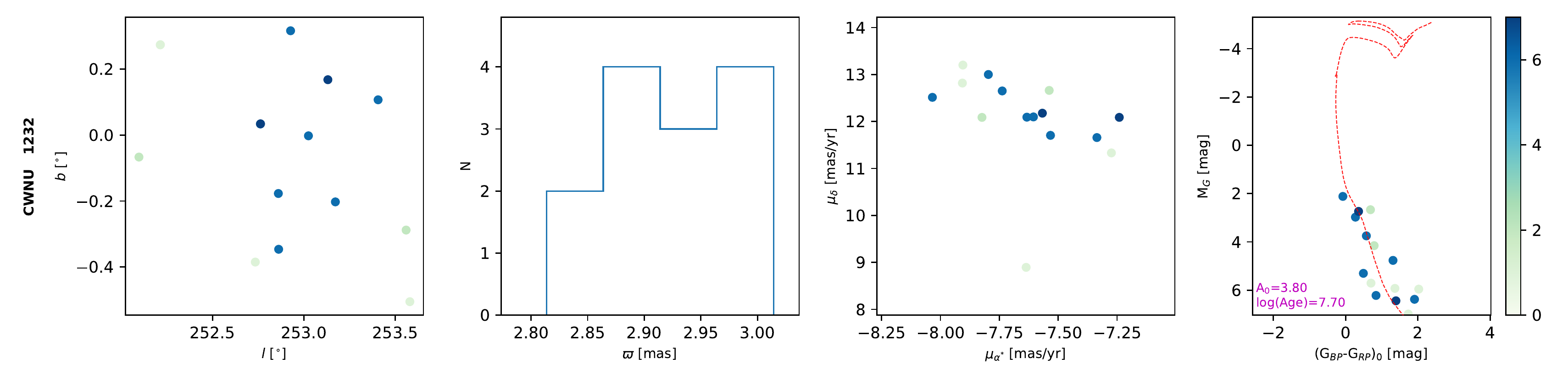}
\includegraphics[width=0.9\linewidth]{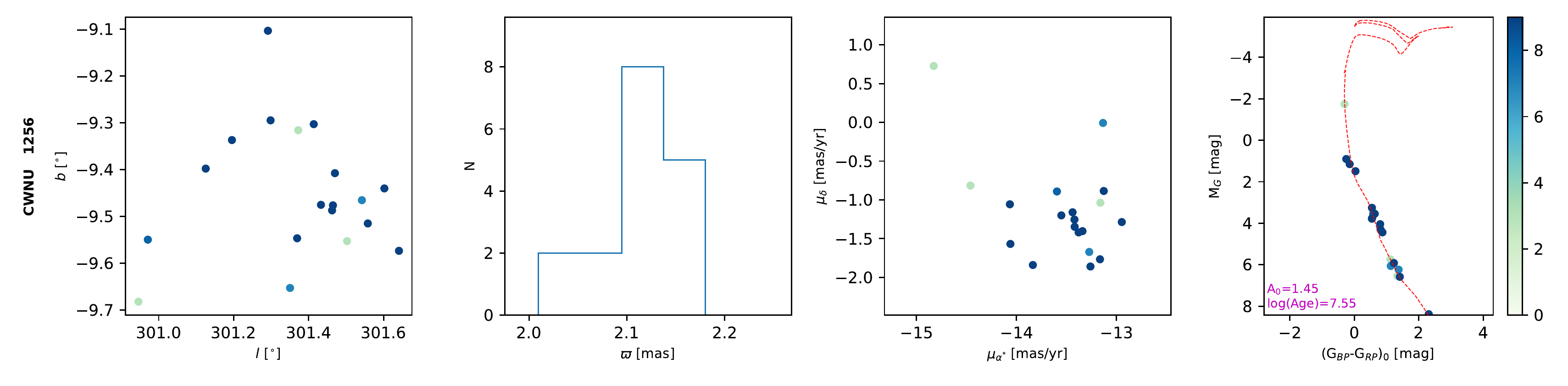}
	\caption{Same as Fig.~\ref{fig_pregaia}, but for the new candidates in class 2 (upper panel) and in class 3 (lower panel), respectively.}
	\label{fig_class2}
\end{center}
\end{figure*}

For the newly found cluster candidates, we visually inspected their isochrone fits and manually classified them according to their CMDs. Our classification principles were similar to those adopted by~\citet{Castro20} and H22: as shown in Fig.~\ref{fig_class1}, the class 1 candidates present clear CMDs and no obvious gap at low mass stars sequence. The rest candidates were classified as class 2 with unclear isochrone fitting, and class 3 with loosely CMD distribution (Fig.~\ref{fig_class2}). The class 1 cluster candidates were labeled as CWNU 1001 - CWNU 1214 while the class 2  and class 3 cluster candidates were labeled as CWNU 1215 - CWNU 1253, and CWNU 1254 - CWNU 1270, respectively. The complete figure sets are available in the online journal.  
 
%
%
 
%
%
%
%
 
\subsubsection{Ages and line-of-sight extinction }\label{sec:statistic}
\begin{figure*}
\begin{center}
	\includegraphics[width=0.9\linewidth]{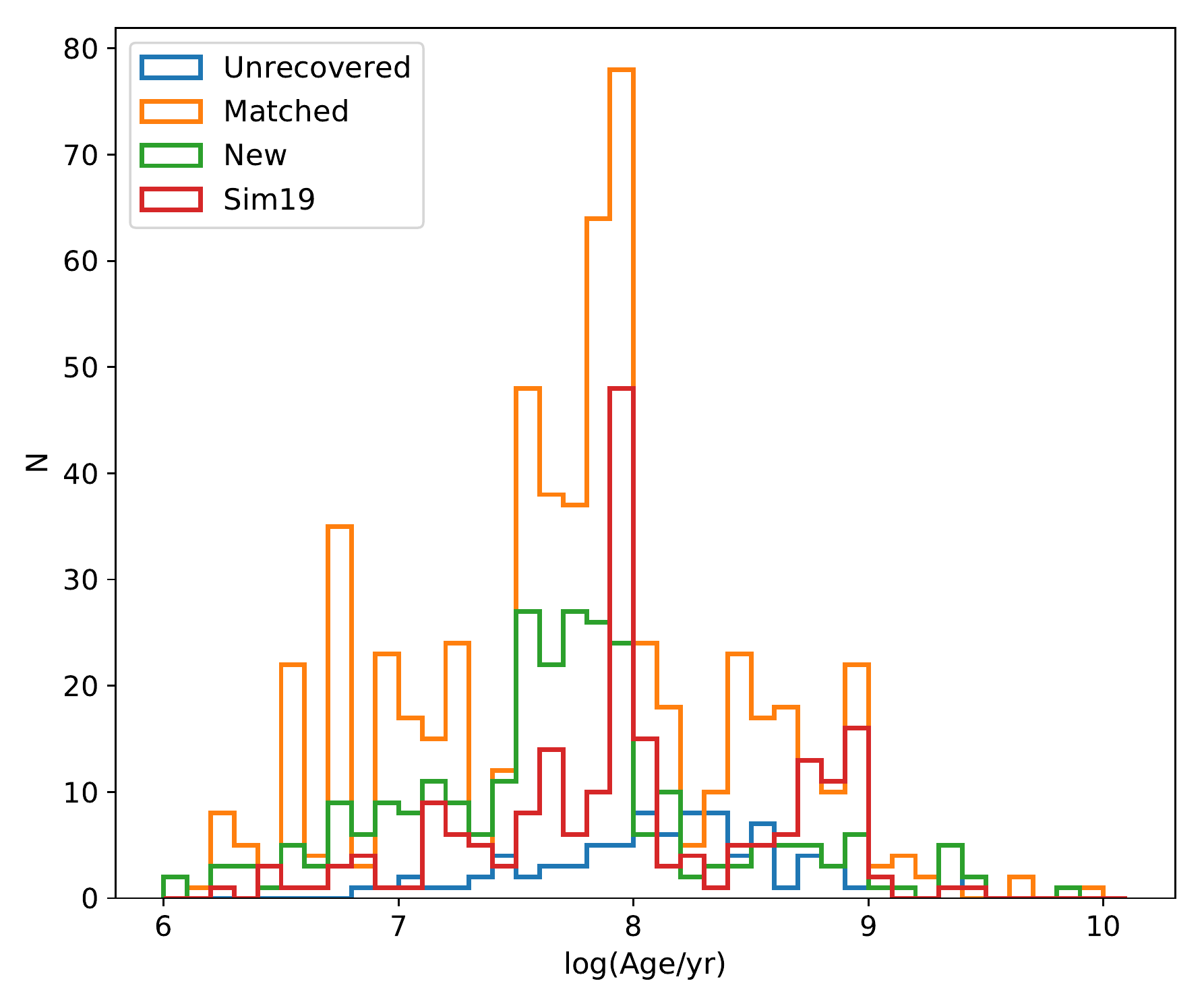}
	\caption{Histograms of the cluster ages derived from isochrone fittings in this work, and cluster ages from~\citet{Sim19}, ~\bf{the OC ages not recovered in this work are from CG20}.}
	\label{fig_age}
\end{center}
\end{figure*}

\begin{figure*}
\begin{center}
	\includegraphics[width=0.9\linewidth]{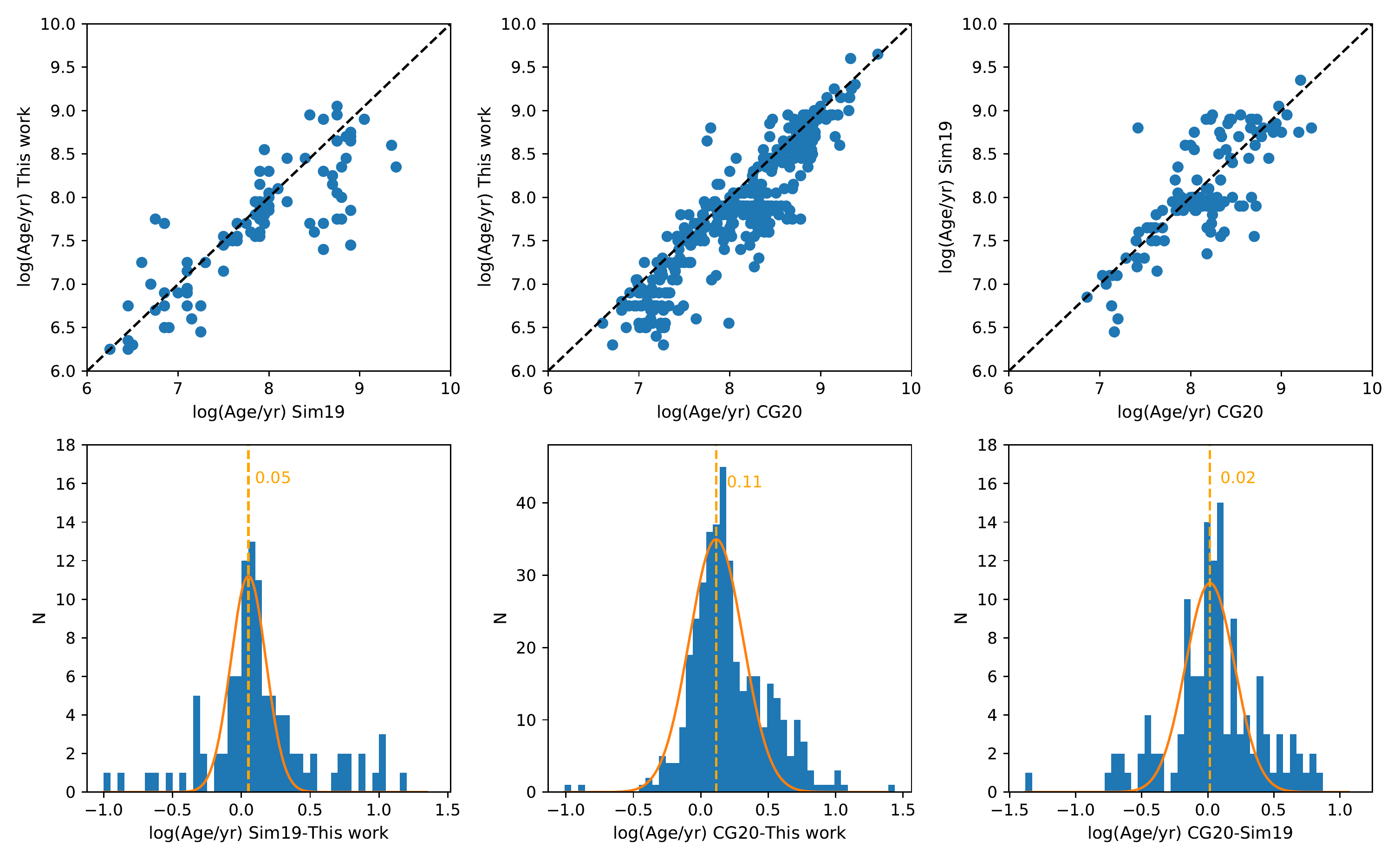}
	\caption{Comparison with cluster ages from this work, from ~\citet{Sim19}, and from CG20. Upper panels: comparison of the ages for the matched clusters. Lower panels: age differences recorded in different catalogs, the orange curves shown the Gaussian fitted distributions.}
	\label{fig_age2}
\end{center}
\end{figure*}

Fig.~\ref{fig_age} shows age histograms of the clusters. The orange and green lines show the ages inferred in this work for the matched clusters and new clusters, respectively, while the red line show the results from~\citet[][hereafter Sim19]{Sim19}. It can be seen that the cluster ages derived in this work are consistent with the age distribution of new clusters found in the Sim19. The age distribution produced two peaks in logarithmic age at 7.9 and $\sim$8.7 dex; that is to say, most of the nearby star clusters were young or intermediate-aged clusters. However, our results have a systematic deviation of 0.11~dex from the age in CG20. As shown in Fig.~\ref{fig_age2}, the age distribution of CG20 is larger than our results, reaching twice the step size of our fitting method. This may be caused by the different fitting methods, in which the least-square method was used for our isochrone fitting (Sec.~\ref{sec:isochrone}), while an artistic neural network method was used by CG20.

\begin{figure*}
\begin{center}
	\includegraphics[width=0.9\linewidth]{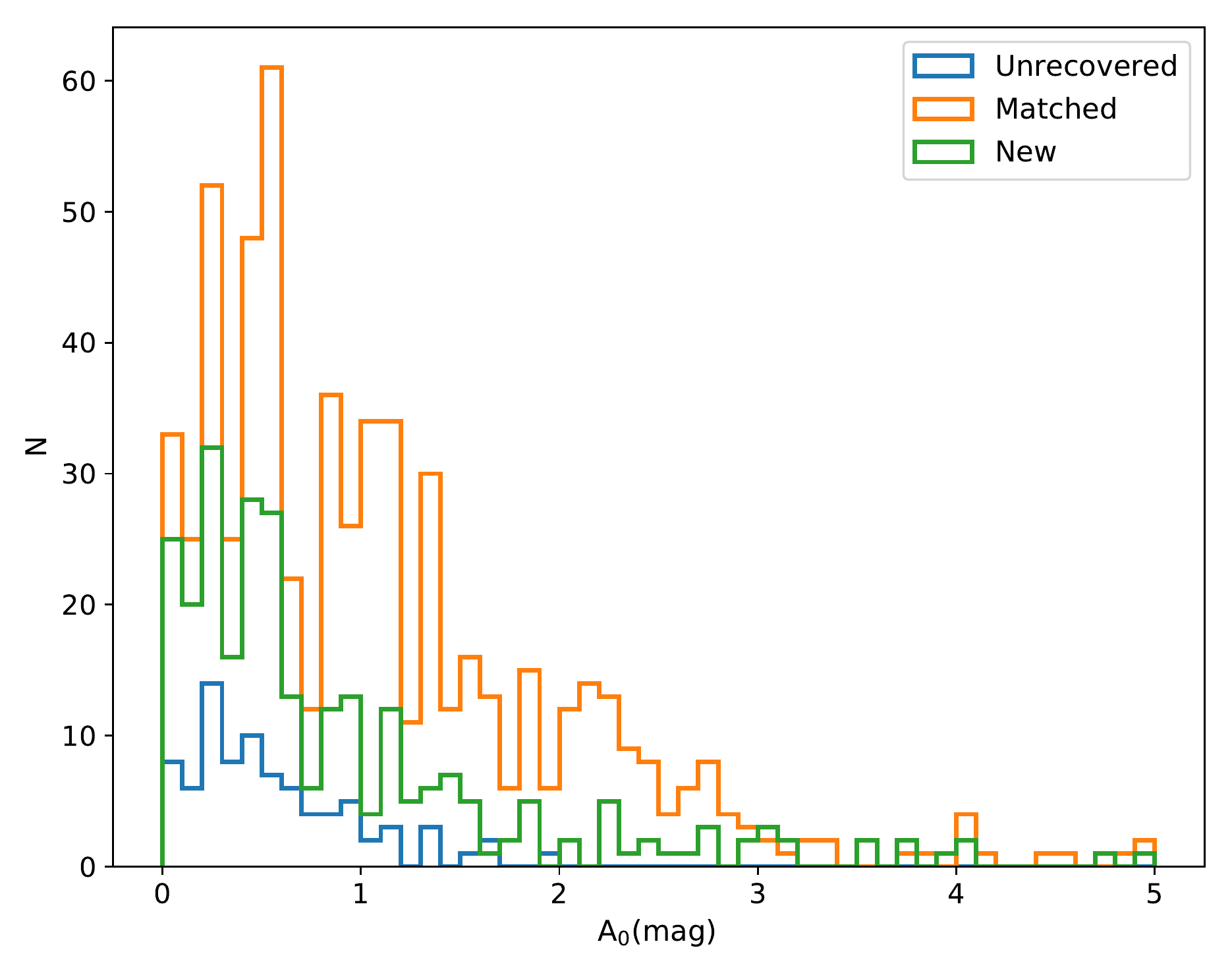}
	\caption{Histograms of the cluster extinctions derived from the isochrone fittings in this work and unrecovered OCs in CG20.}
	\label{fig_ag}
\end{center}
\end{figure*}

\begin{figure*}
\begin{center}
\includegraphics[width=0.9\linewidth]{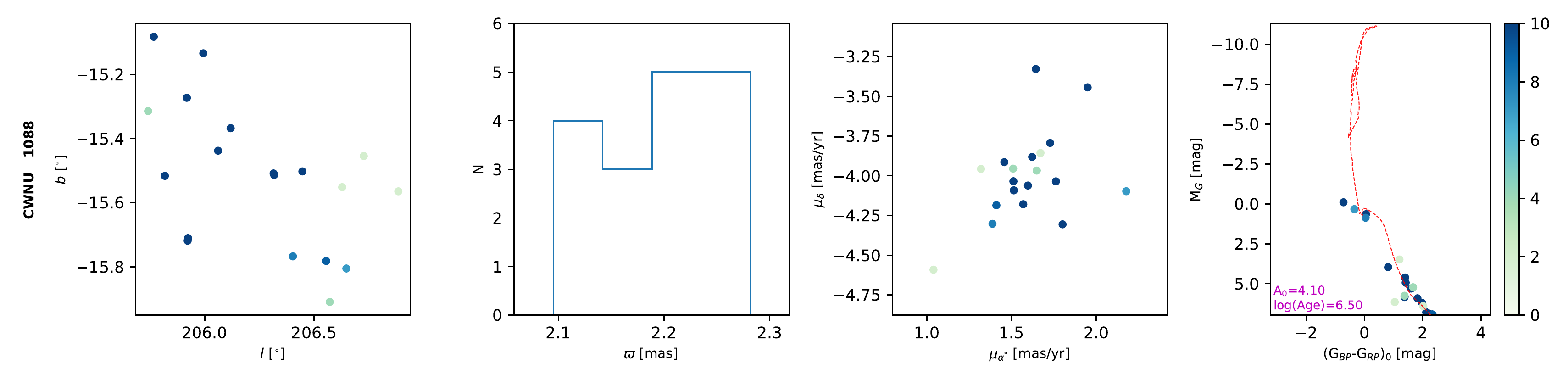}	
\includegraphics[width=0.9\linewidth]{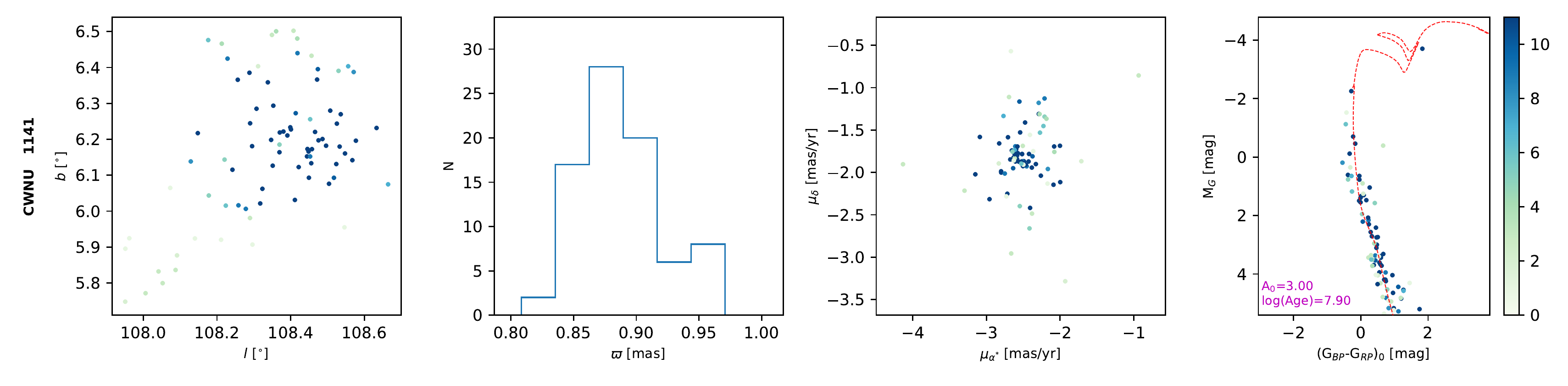}
	\caption{Same as Fig.~\ref{fig_pregaia}, but for the highly extinguished new candidates.}
	\label{fig_high_ag}
\end{center}
\end{figure*}

In Fig.~\ref{fig_ag} we show the resulting extinction histograms, which indicates that the extinction value A$_0$ for most of the nearby star clusters are less than 3~mag, and the known cluster and new candidates had the same distribution. Fig.~\ref{fig_high_ag} shows two star clusters with relatively large extinction. We noted that their foregrounds are star-forming regions with dense dust extinction, e.g., CWNU 1088 and CWNU 1141 were located in/behind the Orion dust ring~\citep{Schlafly15} and Cepheus Flare cloud~\citep{Dame01,Green19}, respectively.

\subsubsection{New candidates at high Galactic latitude}\label{sec:high_b}

\begin{figure*}
\begin{center}
		\includegraphics[width=0.9\linewidth]{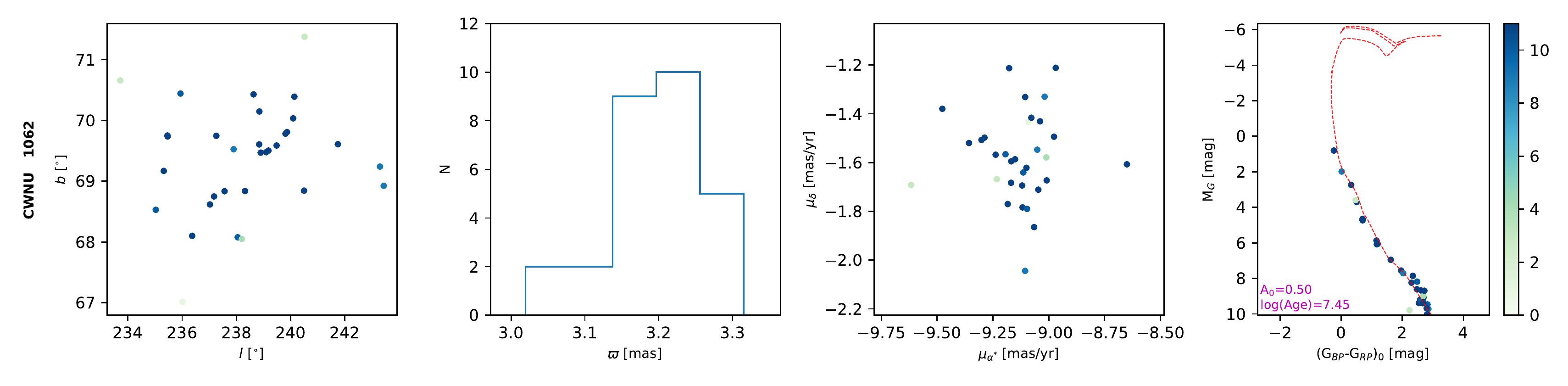}
				\includegraphics[width=0.9\linewidth]{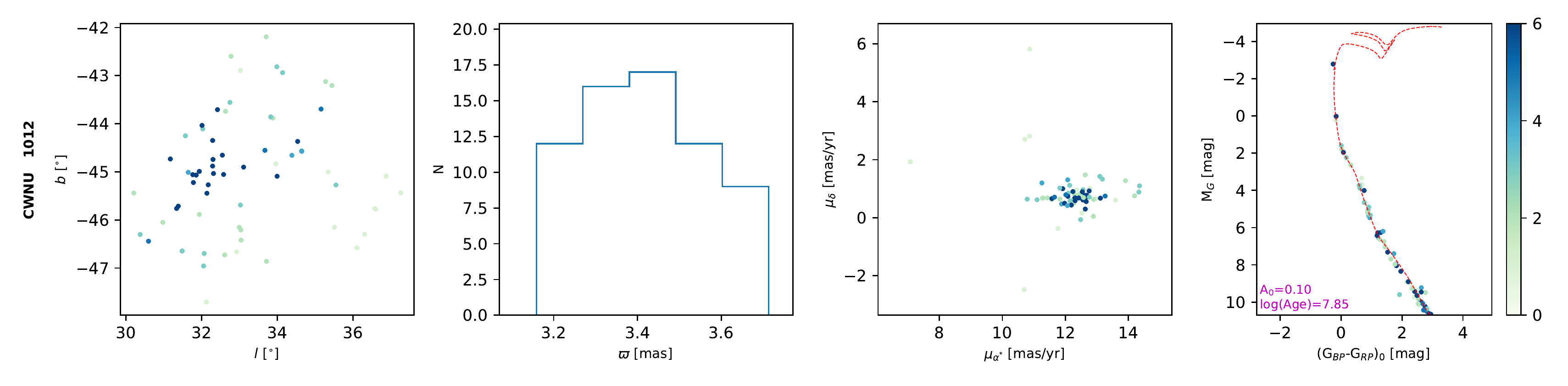}
							\includegraphics[width=0.9\linewidth]{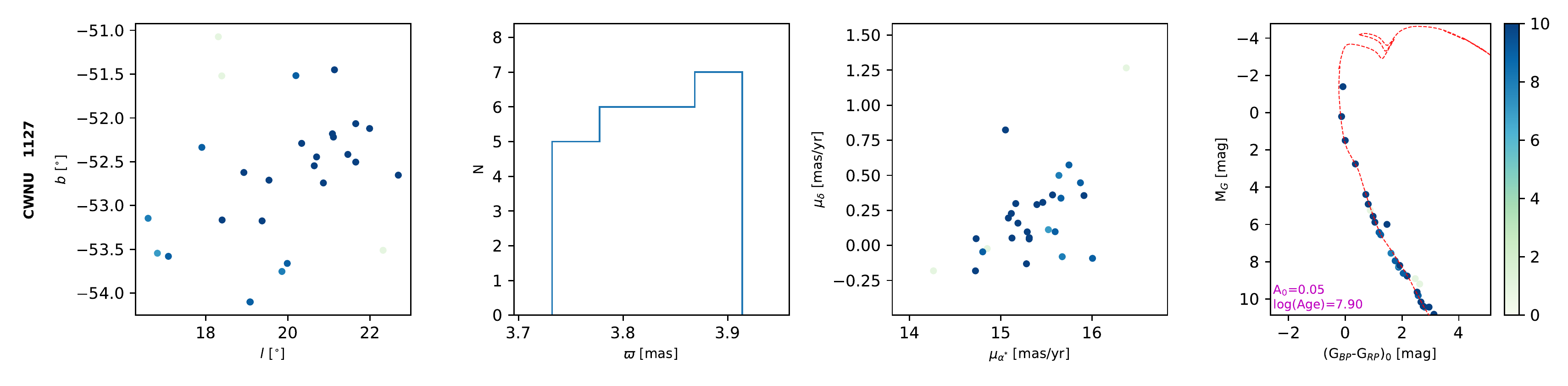}	
	\caption{Same as Fig.~\ref{fig_pregaia}, but for the new candidates at high Galactic latitudes.}
	\label{fig_high_b}
\end{center}
\end{figure*}

Only a few star clusters have been previously found at high Galactic latitudes (in total, 17 clusters with $|b|$ > 20$^\circ$ were cataloged in CG20). In this study, we have detected 46 new cluster candidates with |b| greater than 20 degrees. Some examples are given in Fig.~\ref{fig_high_b}. Most of these candidates are located in low extinction regions and possess clearly defined CMDs. These candidates have wide projections in Galactic coordinates, which is one of the reasons why they were difficult to find in general clustering searches.

\begin{figure*}
\begin{center}
	\includegraphics[width=0.9\linewidth]{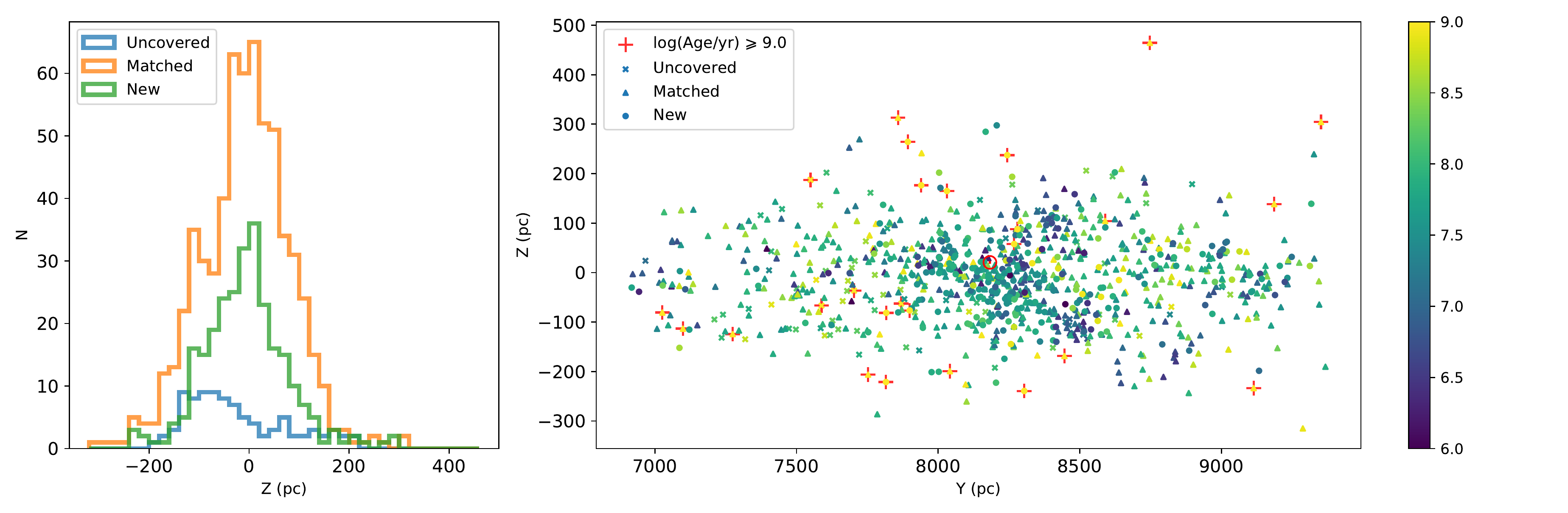}
	\caption{Left: histograms of the clusters' Galactic altitudes. Right: an edge-on view of the clusters, where the color bar presents the logarithmic age of each cluster. The cross symbols indicate the old clusters with age $\geqslant$ 10$^9$ yr.}
	\label{fig_z2}
\end{center}
\end{figure*}

\begin{figure*}
\begin{center}
	\includegraphics[width=0.8\linewidth]{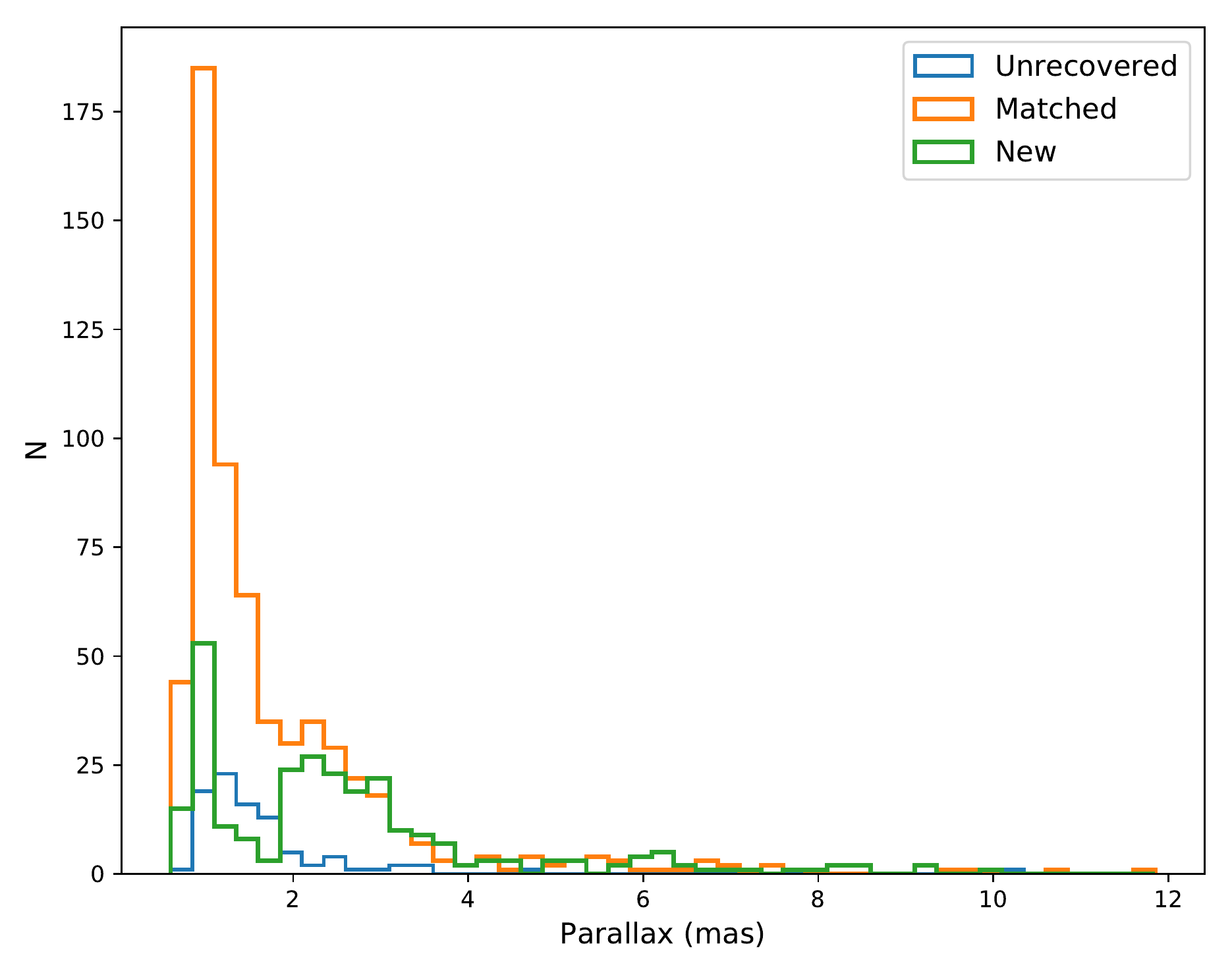}
	\caption{Histograms of the parallax for the new candidates, matched known clusters, and unrecovered CG20 OCs.}
	\label{fig_parallax}
\end{center}
\end{figure*}

In Fig.~\ref{fig_z2}, we have presented Galactic altitude, Z, and Galactocentric radius as a function of age, and most of the young clusters were located no further than 120 pc from the Galactic disk. Meanwhile, for the older clusters, the |Z| values extended to 200 - 400 pc. However, as described above, most of these star clusters at high Galactic latitudes were absent in previous studies. Therefore, as the parallax histograms shown in Fig.~\ref{fig_parallax}, the newly discovered star cluster candidates had significantly increased the proportion of star clusters within $\sim$500 pc. In comparison, the proportion of new candidates between 1.5 to 2 mas was much smaller than the matched known clusters, and most of the unrecovered clusters are also located in the southern Galactic plane, with the distances beyond $\sim$500~pc.

\begin{figure*}
\begin{center}
	\includegraphics[width=1.0\linewidth]{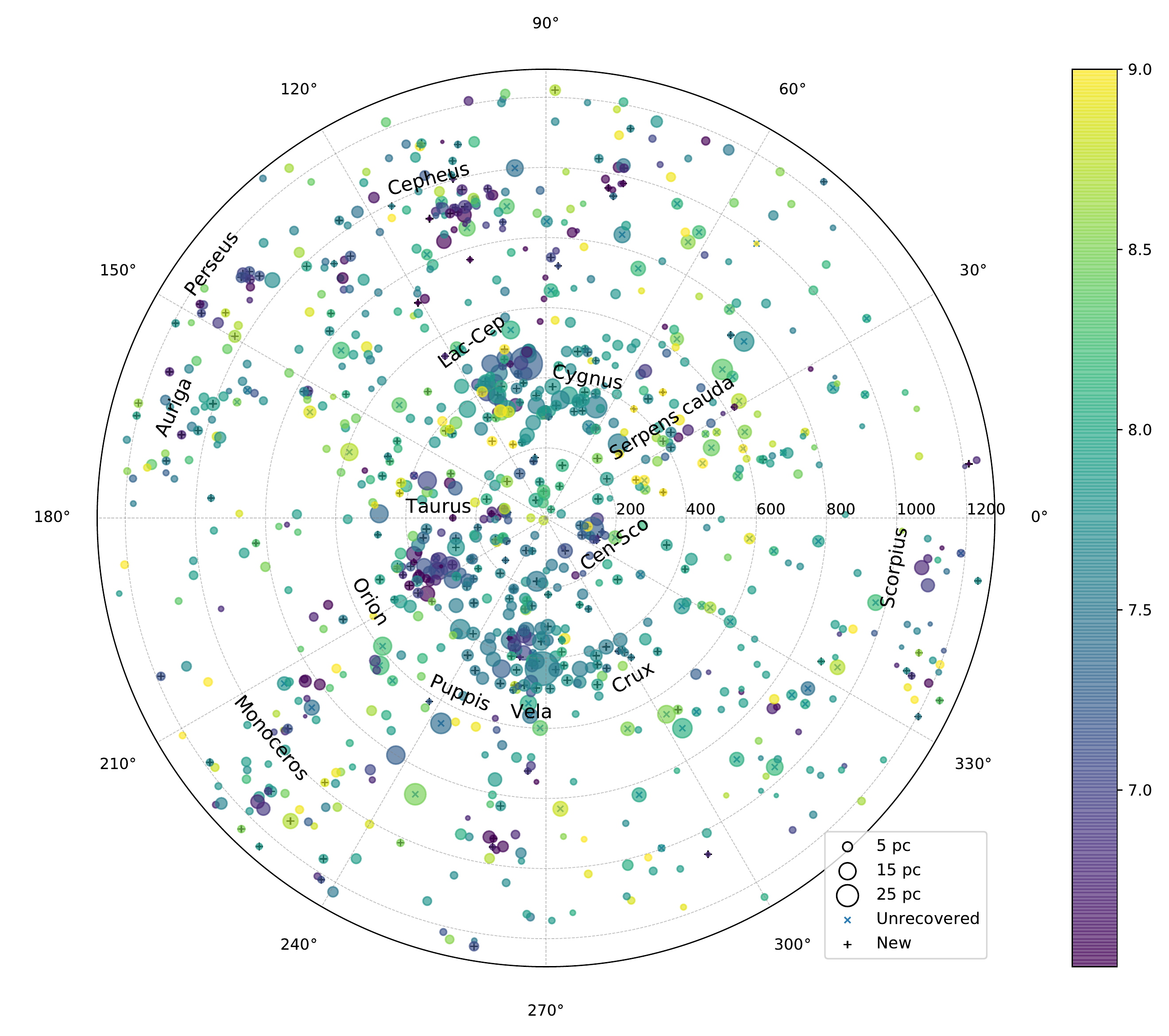}
	\caption{A face-on view of star clusters near the Solar System, the polar coordinates shown the Galactic longitude and the projected distance from the sun (in unit of ~pc). The filled circles present the locations for the new candidates (+), matched clusters, and unrecovered CG20 OCs ($\times$), the logarithmic ages are shown in different colors, and the size of the circles are proportional to the 1~$\sigma$ dispersion (estimated as $\pi/0.18\cdot\varpi^{-1}\cdot\sqrt{\sigma_{l}^2 \cdot cos^2(b)+\sigma_{b}^2}$) of the cluster members.}
	\label{fig_plane}
\end{center}
\end{figure*}

Another reason for the new detection in nearby spaces is that there are a large number of star forming regions near the Solar System. As our investigated area shown in Fig.~\ref{fig_plane}, it can be seen that many young star clusters are located within $\sim$500~pc from the sun. The labels in the figure indicate the direction in which the clusters are more concentrated in Galactic plane, which are also consistent with the regions where the nearby pre-main-sequence stars~\citep{Zari18} are located. The extended large-scale clusters also located in Vela, Orion, and Lac-Cep star forming regions, and their ages are generally not more than $\sim$50 Myr.

\section{Conclusions}\label{sec:summary}

Based on the stellar astrometry in Gaia EDR3, we presented the results of an all-sky search for star clusters, obtaining 886 nearby clusters and candidates. We cross-matched them with published clusters in Gaia DR2 and found 616 matched clusters. Additionally, we identified 270 new cluster candidates, with classifications based on CMDs and isochrone fittings. We provided a catalog of positions, proper motions, and parallaxes for all clusters, and physical parameters including age and extinction, as well as membership for each cluster. 

We were able to determine more details for a number of the matched clusters relative to previous studies. Most of the new memberships had more stars than shown in the literature. The richer halo profiles and extended structures of some star clusters were well presented. Compared with previous studies, we found more cluster samples within 500 pc of the Solar System, especially in regions with Galactic latitudes greater than 20 degrees. Most of new cluster candidates had small extinction values, but some clusters behind a foreground of dust clouds had larger extinction values. 

Our study has shown that current Gaia data were still not fully utilized in the search for clusters, and there were many unknown stellar aggregates that had not been discovered by previous investigations. At the same time, Gaia DR3 contained four times more stellar radial velocity values than found in Gaia DR2, which will also greatly help us to better study the kinematic characteristics of the adjacent space of the Solar System. 

\section{Acknowledgements}
We thank the reviewer for the constructive comments, and we thank Dr. Li Chen and Dr. Jing Zhong for reading this article and putting forward useful suggestions. 
This work has made use of data from the European Space Agency (ESA) mission GAIA (\url{https://www.cosmos.esa.int/gaia}), processed by the GAIA Data Processing and Analysis Consortium (DPAC,\url{https://www.cosmos.esa.int/web/gaia/dpac/consortium}). Funding for the DPAC has been provided by national institutions, in particular the institutions participating in the GAIA Multilateral Agreement.
This work is supported by Fundamental Research Funds of China West Normal University (CWNU, No.21E030), the Innovation Team Funds of CWNU, the National Natural Science Foundation of China (NSFC, No.12003022). We acknowledge the science research grants from the Chinese Space Station Telescope project with NO. CMS-CSST-2021-B03.
L.Y.P is supported by the National Key Basic R\&D Program of China via 2021YFA1600401, the National Natural Science Foundation of China (NSFC) under grant 12173028, the CSST project: CMS-CSST-2021-A10, the Sichuan Science and Technology Program (Grant No. 2020YFSY0034), the Sichuan Youth Science and Technology Innovation Research Team (Grant No. 21CXTD0038),and the Innovation Team Funds of CWNU(Grant No. KCXTD2022-6).

\bibliographystyle{aasjournal} 
\bibliography{oc} 


\end{CJK*}

\end{document}